\documentclass[twocolumn,10pt,aps,pra,showpacs,longbibliography,superscriptaddress]{revtex4-1}

\usepackage{amsmath,amssymb,mathtools,xcolor}
\usepackage{graphicx}
\usepackage[T1]{fontenc}
\usepackage{hyperref}
\usepackage[utf8]{inputenc}

\inputencoding{utf8}

\begin{document}

\title{Laughlin-like states in bosonic and fermionic atomic synthetic ladders}

\author{Marcello~Calvanese~Strinati}
\affiliation{NEST, Scuola Normale Superiore and Istituto Nanoscienze-CNR, I-56126 Pisa, Italy}
\affiliation{D\'epartement de Physique, Ecole Normale Sup\'erieure / PSL Research University,
CNRS, 24 rue Lhomond, F-75005 Paris, France}

\author{Eyal~Cornfeld}
\affiliation{Raymond and Beverly Sackler School of Physics and Astronomy, Tel-Aviv University, IL-69978 Tel Aviv, Israel}

\author{Davide~Rossini}
\affiliation{NEST, Scuola Normale Superiore and Istituto Nanoscienze-CNR, I-56126 Pisa, Italy}
\affiliation{{Dipartimento di Fisica, Universit\`a di Pisa and INFN, Largo Pontecorvo 3, I-56127 Pisa, Italy}}

\author{Simone~Barbarino}
\affiliation{NEST, Scuola Normale Superiore and Istituto Nanoscienze-CNR, I-56126 Pisa, Italy}
\affiliation{SISSA-International School for Advanced Studies, Via Bonomea 265, I-34136 Trieste, Italy}

\author{Marcello~Dalmonte}
\affiliation{ICTP, Strada Costiera 11, 34151 Trieste, Italy}
\affiliation{Institute for Quantum Optics and Quantum Information of the Austrian Academy of Sciences, 6020 Innsbruck, Austria}
\affiliation{Institute  for  Theoretical  Physics,  University  of  Innsbruck,  6020  Innsbruck,  Austria}

\author{Rosario~Fazio}
\affiliation{ICTP, Strada Costiera 11, 34151 Trieste, Italy}
\affiliation{NEST, Scuola Normale Superiore and Istituto Nanoscienze-CNR, I-56126 Pisa, Italy}

\author{Eran~Sela}
\affiliation{Raymond and Beverly Sackler School of Physics and Astronomy, Tel-Aviv University, IL-69978 Tel Aviv, Israel}

\author{Leonardo~Mazza}
\affiliation{D\'epartement de Physique, Ecole Normale Sup\'erieure / PSL Research University,
CNRS, 24 rue Lhomond, F-75005 Paris, France}

\date{\today}

\begin{abstract}
The combination of interactions and static gauge fields plays a pivotal role in our understanding of strongly-correlated quantum matter. Cold atomic gases endowed with a synthetic dimension are emerging as an ideal platform to experimentally address this interplay in quasi-one-dimensional systems. 
A fundamental question is whether these setups can give access to pristine two-dimensional phenomena, such as the fractional quantum Hall effect, and  how. 
We show that unambiguous signatures of bosonic and fermionic Laughlin-like states can be observed and characterized in synthetic ladders. 
We theoretically diagnose these Laughlin-like states focusing on the chiral current flowing in the ladder, on the central charge of the low-energy theory, and on the properties of the entanglement entropy. 
Remarkably, Laughlin-like states are separated from conventional liquids by Lifschitz-type transitions, characterized by sharp discontinuities in the current profiles, which we address using extensive simulations based on matrix-product states. 
Our work provides a qualitative and quantitative guideline towards the observability and understanding of strongly-correlated states of matter in synthetic ladders. In particular, we unveil how state-of-the-art experimental settings constitute an ideal starting point to progressively tackle two-dimensional strongly interacting systems from a ladder viewpoint, opening a new perspective for the observation of non-Abelian states of matter.
\end{abstract}


\maketitle

\section{Introduction}

Topological order is one of the most fascinating discoveries in physics of the last decades. Owing to their highly non-local nature, topological phases can host anyonic excitations, which lie at the heart of several schemes for fault-tolerant quantum computation~\cite{RevModPhys.80.1083}. For this reason, the experimental realization of Abelian and non-Abelian anyons with ultra-cold gases, which offer unprecedented possibilities for the coherent manipulation of quantum systems, has motivated an impressive amount of theoretical studies~\cite{10.1038/nphys2259,10.1038/nphys3803}. In particular, most of the work has focused on the fractional quantum Hall effect (FQHE)~\cite{ezawa2008QHE}, and on topologically-ordered states stemming from the interplay of interactions and static magnetic fields (or more general gauge fields)~\cite{RevModPhys.83.1523}. However, despite of impressive experimental progresses in engineering the latter using either rotating gases, engineered lattice shaking, or laser-assisted tunneling, the regimes of stability of such phases have been so far elusive. 

In this article, we demonstrate that fundamental features of FQH states can be realized and observed in current cold-atom experiments exploiting the concept of \textit{synthetic dimension}~\cite{PhysRevLett.108.133001,PhysRevLett.112.043001,PhysRevLett.115.195303,PhysRevA.95.023607}, namely interpreting internal atomic states as spatial indices. Recently, this concept led to the experimental observation of chiral edge states in both bosonic and fermionic quantum gases~\cite{science1510, science1514}. The key experimental advantage of this protocol is that the gauge field is implemented without significant spontaneous emission rates, thus guaranteeing sufficiently long coherence times to observe strongly correlated states of matter (especially when clock transitions are employed as in Ref.~\cite{PhysRevLett.117.220401}). As such, synthetic ladders represent ideal settings to investigate the interplay between gauge fields and interactions considered in numerous theoretical studies, both for bosons~\cite{PhysRevB.64.144515,PhysRevB.72.104521,PhysRevB.73.100502, PhysRevA.85.041602, PhysRevLett.111.150601, PhysRevB.87.174501, PhysRevA.89.063617, 1367-2630-16-7-073005, PhysRevB.91.054520, DiDio2015, PhysRevB.91.140406, PhysRevA.92.013625, PhysRevB.92.060506, PhysRevB.92.115446, 1367-2630-17-9-092001, PhysRevB.92.115120, PhysRevA.92.053623, PhysRevLett.115.190402, PhysRevA.93.053629, PhysRevA.94.023630, PhysRevA.94.063628,1367-2630-18-5-055017,PhysRevA.94.063632} and for fermions~\cite{PhysRevB.71.161101, PhysRevB.73.195114, PhysRevB.76.195105, 1367-2630-17-10-105001,ncomms9134,PhysRevLett.115.095302,1367-2630-18-3-035010,arxiv1610.00281,PhysRevB.92.115446,arXiv:1607.07842,PhysRevA.93.013604,PhysRevA.93.023608}. 

\begin{figure}[t]
\centering
\includegraphics[width=8cm]{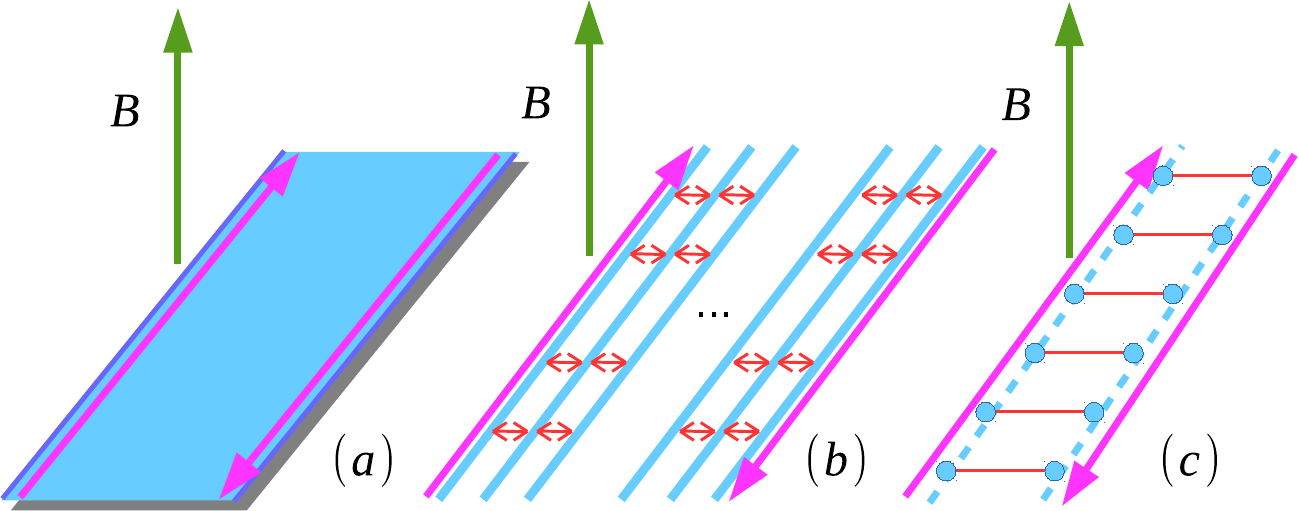}
\caption{(Color online) From Laughlin to Laughlin-like states. (a) A two-dimensional FQH system with counter-propagating fractional edge modes (purple arrows) is split into (b) a long array of  one-dimensional wires, where the edge modes localize on the extremal ones. Finally, (c) Laughlin-like physics is studied in the limit of two coupled chains (a ladder).}
\label{fig:schemekane}
\end{figure}

We show that the one-dimensional version of the Laughlin state appears in several experimentally-relevant models of one-dimensional quantum gases with a synthetic dimension.
We illustrate that unambiguous signatures of this Laughlin physics can be characterized  via measurements of the current {circulating along the system}, thus drawing a clear path towards the experimental detection of FQHE in {cold-atom setups}.

The rationale behind this can be understood as follows (see Fig.~\ref{fig:schemekane}): a genuine Laughlin state in a two-dimensional system with hard-wall boundaries is characterized by a chiral edge mode with fractional conductance. Influenced by the coupled-wire approach to the FQHE proposed in Refs.~\cite{PhysRevLett.88.036401,PhysRevB.89.085101}, we split the two-dimensional system into a long array of one-dimensional wires pierced by a magnetic flux, and subsequently reduce the length of the array to two neighboring and coupled wires~\cite{PhysRevB.89.115402, PhysRevB.92.115446}. During this procedure, the Laughlin state has been turned into a one-dimensional quantum liquid with two counter-propagating fractional modes, each of them localized at one of the two wires. This is what we call a \textit{Laughlin-like state}, and it is the object of our analysis.

By interpreting the quantum number labelling the wires as the spin state of an atom, we obtain a simple model for a one-dimensional quantum gas with synthetic dimension, or, briefly, the synthetic ladder mentioned above that is currently at experimental focus. A similar analysis can also be done for condensed-matter systems like Rashba wires~\cite{PhysRevB.89.115402}, which are unfortunately prone to disorder and to localization of the counter-propagating modes: ultra-cold atoms are disorder-free and thus provide a more convenient platform for an experimental analysis.

We provide numerical evidence of Laughlin-like states using algorithms based on matrix-product states (MPS)~\cite{Schollwock201196}.
With the help of analytical techniques such as bosonization~\cite{giamarchi2003quantum,gogolin2004bosonization}, we pinpoint some unambiguous signatures of Laughlin-like physics, and characterize them with our simulations. 
In particular, we focus on the singular behavior of the chiral current flowing in the system when entering the Laughlin-like state and on entanglement-related observables.
An exactly-solvable model~\cite{PhysRevB.84.085434,PhysRevB.92.115446,arXiv:1607.07842} is {also} presented which nicely complements the approximate analytical techniques and the numerical simulations.

Our work is of direct experimental relevance for laboratories where gases with a synthetic dimension are currently realized.
The experimental observation of states with counter-propagating fractional modes would constitute the first unambiguous signature of effects quintessential to topological strongly-correlated systems. Synthetic dimensions thus mark a new paradigm in the quantum simulation of the FQHE and this study opens the way to the discussion of topological order and anyons in systems with a longer synthetic dimension.

The article is organized as follows: 
in Sec.~\ref{laughlinstatesinbosonicmodels}, we present our numerical results on the observation of Laughlin-like states in bosonic models, whereas in Sec.~\ref{sec:laughlinstatesinfermionicmodels} we focus on fermionic systems {and} discuss the exactly-solvable model which has been anticipated above. In the following two sections, we critically discuss the signatures employed to diagnose the Laughlin-like physics. In particular, in Sec.~\ref{sec:IQHE}, {we compare our results with those of a non-interacting fermionic ladder displaying integer quantum Hall effect-like physics.} In Sec.~\ref{sec:signaturesbasedonthecurrent}, we present two arguments supporting the {fact that the singular behavior of the current, which we observe in our numerics, is related to Laughlin-like physics}. {We discuss the experimental realizability of our setup in Sec.~{VI}}. Conclusions are drawn in Sec.~\ref{sec:conclusions}. Throughout the paper, we set $\hbar=1$ and the lattice constant $a=1$.

\section{Laughlin-like states in bosonic ladders}
\label{laughlinstatesinbosonicmodels}
Although Laughlin states are usually studied in the two-dimensional FQHE, {their signatures} also appear in strongly anisotropic systems.
Following the analysis in arrays of many coupled wires~\cite{PhysRevLett.88.036401,PhysRevB.89.085101},
a minimal system consisting of a ladder with the number of legs equal to two and pierced by a magnetic field is predicted to support Laughlin-like states which share important properties with their original version~\cite{PhysRevB.89.115402, PhysRevB.91.054520, PhysRevB.92.115446}.
The goal of this section is the numerical characterization of the fractional Laughlin-like state which appears in an experimentally-relevant microscopic model for a two-leg bosonic ladder.

\begin{figure}[t]
\centering
\includegraphics[width=8cm]{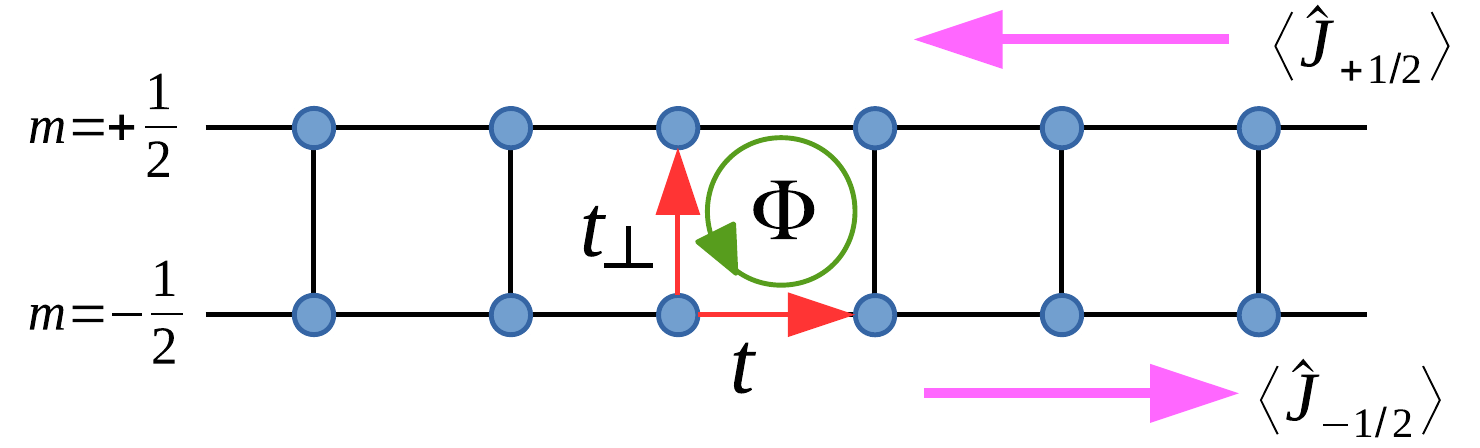}
\caption{(Color online) Scheme of the two-leg ladder (spin projections $m=\pm1/2$) in a magnetic flux, $\Phi$. Longitudinal and transverse hopping parameters are denoted by $t$ and $t_\perp$ respectively. The currents, for the two spin components, $\langle\hat J_{\pm\frac{1}{2}}\rangle$, flow in opposite directions (purple arrows).}
\label{fig:ladderscheme}
\end{figure}

\begin{figure*}[t]
\centering
\includegraphics[width=6.0cm]{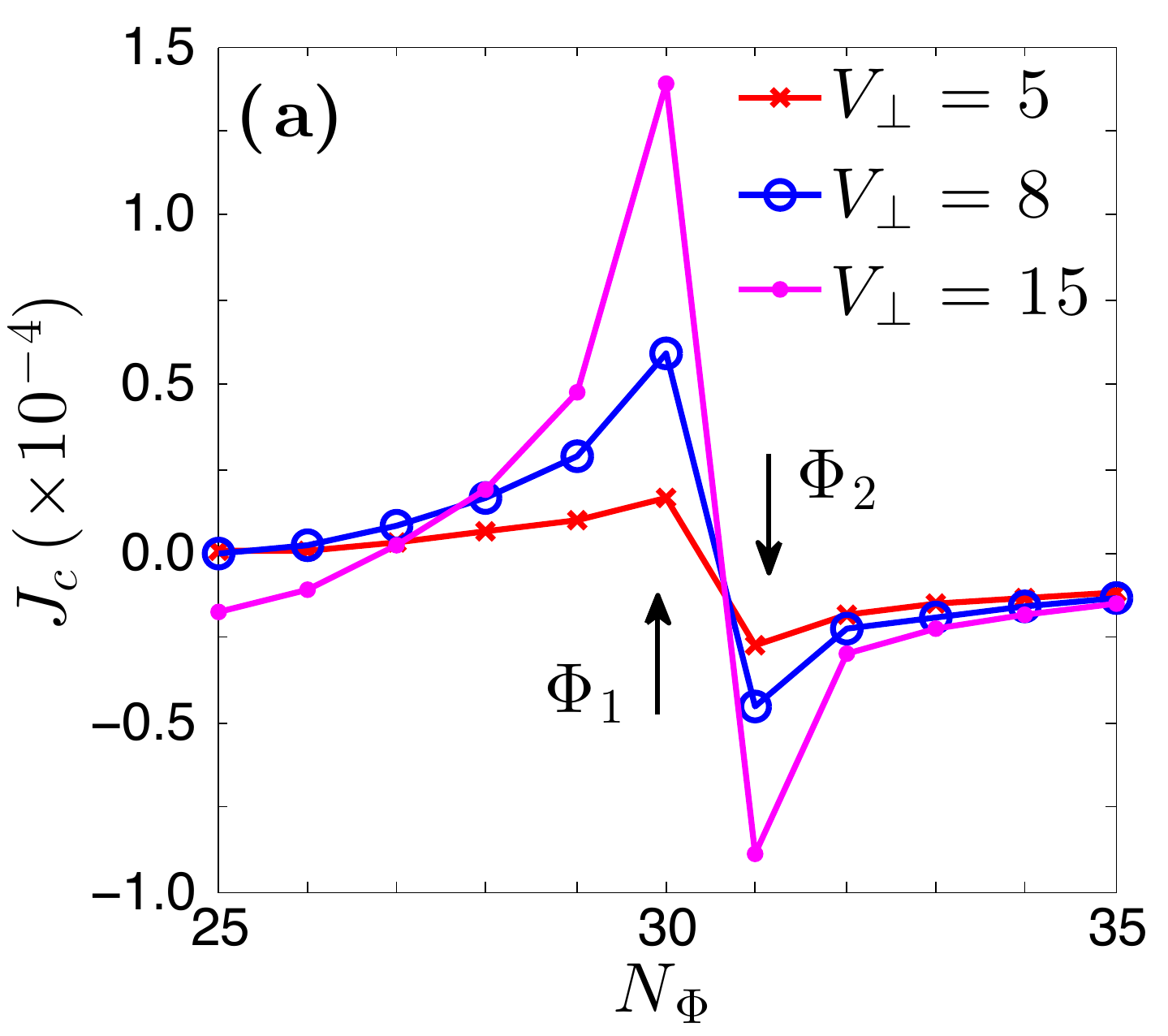}
\includegraphics[width=5.5cm]{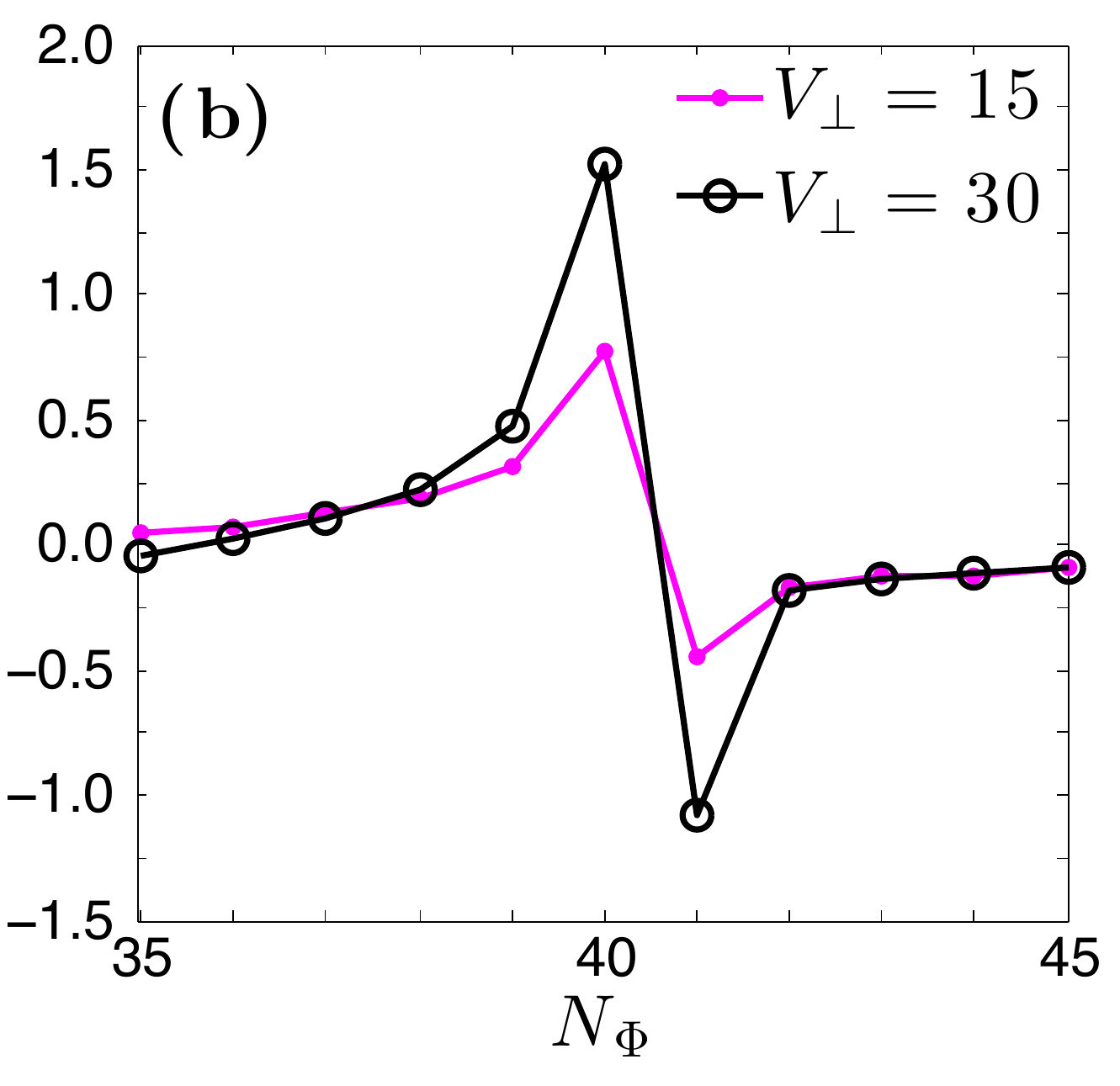}
\includegraphics[width=5.6cm]{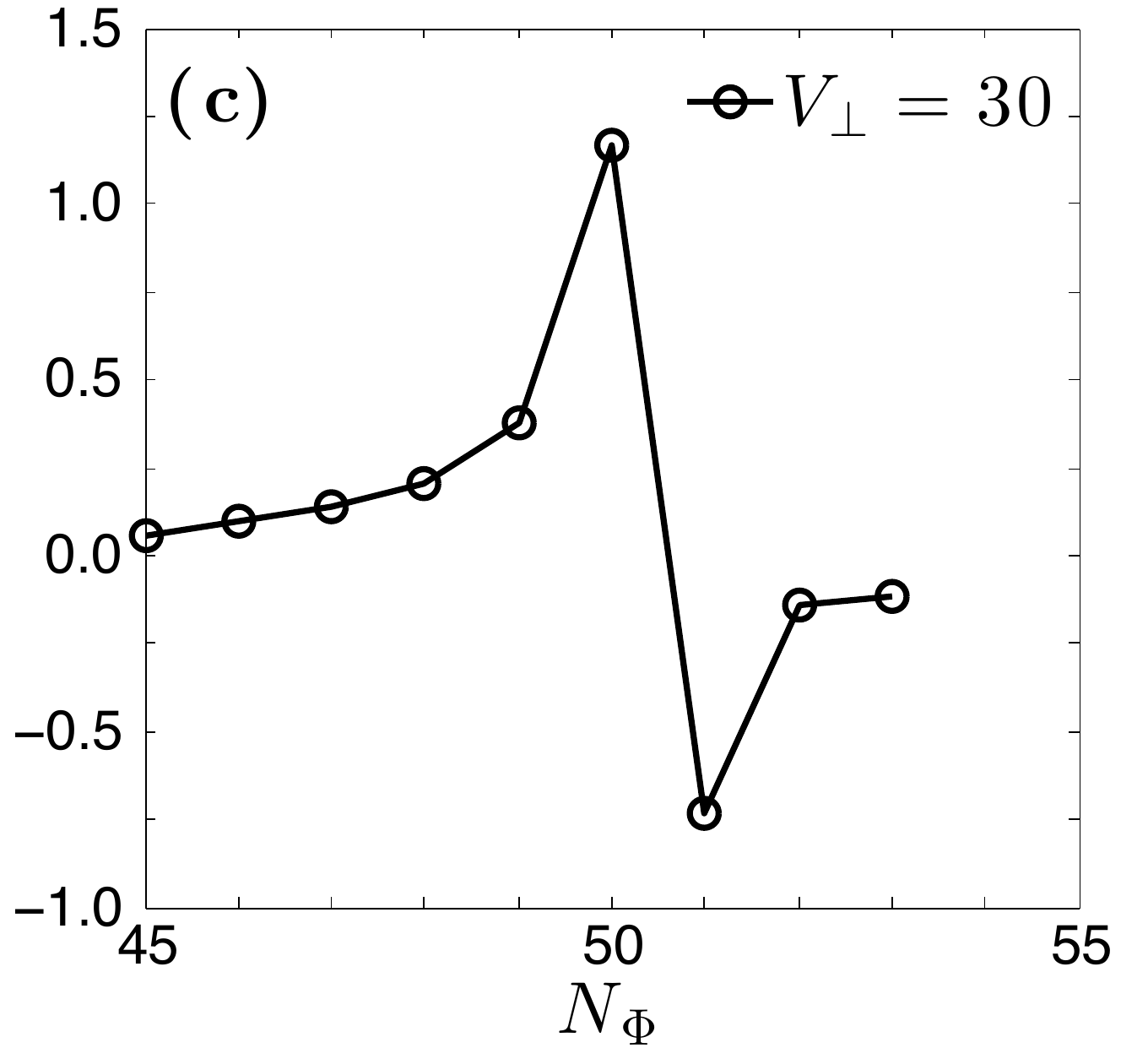}
\caption{(Color online) Numerical results for $J_c$ (in units of $t$) of the bosonic $\nu=1/2$ Laughlin-like state versus flux. The flux is parametrized as $\Phi=2\pi N_\Phi/(L+1)$. Simulation parameters: $L=120$, $D_{\rm max}=180$, $t_\perp/t=10^{-2}$ and $N=30,40,50$ for panels \textbf{(a)}, \textbf{(b)} and \textbf{(c)} respectively. Several values of $V_\perp/t$ are considered. The resonance $\Phi=2\pi n$ for the $\nu=1/2$ Laughlin-like state appears at $N_\Phi = N$. }
\label{fig:numericalsimulationsbosons}
\end{figure*}

{The appearance of Laughlin-like states is linked to a fractional filling factor, relating the particle density to the magnetic flux. They are} identified by the existence of two counter-propagating modes, each of them localized at one of the two legs of the ladder. {Since they are arranged next to each other}, the low-energy physics is not anymore topologically protected (in the two-dimensional FQHE they are crucially separated by a macroscopic distance), and thus we cannot speak of genuine topological order. However, {in the absence of disorder}, the low-energy physics is completely analogous, and in particular the edge modes carry a fractional charge and have a quantized fractional conductance.
Thus, by identifying the two legs with two spin states, as in a synthetic ladder, we can identify the Laughlin-like states with fractional helical liquids~\cite{PhysRevB.89.115402}.

\subsection{Synthetic ladder of hard-core bosons}

{In this section, we discuss Laughin-like states in bosonic synthetic ladders pierced by a flux. The experimental realization of such ladders has been reported in Ref.~\cite{science1514} using $^{87}$Rb atoms in an optical lattice (we also mention Ref.~\cite{NatPhys10.588.14}, where the flux ladder is realized without using a synthetic dimension).}

{Our study is performed in the limit of hard-core bosons (HCB), when the on-site repulsion is so strong to forbid double occupancies in each bosonic site.} We model the synthetic bosonic ladder with the following Hamiltonian: 
\begin{align}
\hat H =&\sum_{j} \Big[ -t\sum_{m=\pm1/2}\left(\hat b^\dag_{j,m}\hat b_{j+1,m}+{\rm H.c.}\right) \nonumber\\
& +t_\perp \left(e^{-i\Phi j}\,\hat b^\dag_{j,-\frac{1}{2}}\hat b_{j,+\frac{1}{2}}+{\rm H.c.}\right)+V_\perp \hat n_{j,+\frac{1}{2}}\hat n_{j,-\frac{1}{2}}\,\, \Big] \,\, .
\label{eq:hamiltonianhardcorebosons}
\end{align}
Here, $t$ and $t_\perp$ are the longitudinal and transverse hopping parameters, $\Phi$ is the flux per plaquette of the orbital magnetic field piercing the ladder, $V_\perp$ represents the on-site density-density interaction between the two legs, and $\hat n_{j,m}=\hat b^\dag_{j,m}\hat b_{j,m}$. Here, $\hat b_{j,m}$ ($\hat b^\dag_{j,m}$) represents the HCB annihilation (creation) operator, which satisfies mixed commutation relation~\cite{prog.theor.phys.16.569}, i.e. $\{\hat b_{j,m},\hat b^\dag_{j,m}\}=1$ and $[\hat b_{j,m},\hat b^\dag_{j',m'}]=0$ for $j\neq j'$ and/or $m\neq m'$, and thus can be represented by spin-1/2 operators. {If not explicit}, we use $t$ as a reference energy scale.

A sketch of the two-leg ladder is shown in Fig.~\ref{fig:ladderscheme}. {We denote by $L$ the number of rungs (which corresponds to the number of minima of the optical lattice potential), and the number of bosons by $N$. We then define} the density $n = N/L$. A crucial quantity in the study of Laughlin-like physics is the filling factor $\nu=\pi n/\Phi$, which relates the density of bosons to the density of magnetic fluxes and parallels the filling factor which is crucial in the study of the two-dimensional FQHE.

Bosonization predicts the appearance of Laughlin-like states in bosonic ladders when the filling factor takes the values $\nu=1/p$, and $p$ is a \textit{even} positive integer~\cite{PhysRevLett.88.036401,PhysRevB.89.085101,PhysRevB.92.115446,PhysRevB.91.054520}; in the following, we focus on the Laughlin-like state appearing at $\nu = 1/2$, which is the most easily accessible state. We stress that bosonization predictions are obtained in the perturbative regime $t_\perp \ll t$ (see Appendix~\ref{sec:bosonizationconventions}), and thus in this article we will restrict our analysis to it in order to have a theoretical guideline for our simulations. According to bosonization, distinctive signatures of the Laughlin-like regime are provided by transport measurements (e.g.\ the fractional conductance)~\cite{PhysRevB.89.115402,PhysRevB.89.045111,PhysRevB.91.115427}. However, the occurrence of the $\nu=1/2$ state can also be detected by thermodynamic observables more suitable for a cold-atom experiment, such as the anisotropy of spin susceptibilities~\cite{PhysRevB.88.035437} and the chiral current, $J_c$~\cite{PhysRevB.92.115446}; we focus on the latter, because it is immediately accessible both in experiments and in numerical simulations.

In the two-leg ladder, the chiral current is proportional to the difference between the currents circulating along the two legs (see Fig.~\ref{fig:ladderscheme}), i.e., $J_c=\frac{1}{2L}\sum_j(\langle\hat J_{j,+\frac{1}{2}}\rangle-\langle \hat J_{j,-\frac{1}{2}}\rangle) =\frac{1}{2}(\langle \hat J_{+ \frac 12} \rangle - \langle \hat J_{- \frac 12} \rangle)$, where the spin-resolved current operator on the link between site $j$ and $j+1$ is
\begin{equation}
\hat J_{j,m}=-it\left(\hat b^\dag_{j,m}\hat b_{j+1,m}-{\rm H.c.}\right) \,\, .
\label{eq:spinresolvedcurrentoperator}
\end{equation}
For symmetry reasons, $\langle\hat J_{+\frac{1}{2}}\rangle=-\langle\hat J_{-\frac{1}{2}}\rangle$. As we will analytically show in Secs.~\ref{sec:IQHE} and~\ref{sec:signaturesbasedonthecurrent}, in the limit $t_\perp/t\ll1$, the chiral current is expected to display a characteristic double-cusp pattern in proximity of the resonant value $\Phi_0 =2\pi n$. 
Close to $\Phi_0$, there is a flux value $\Phi_1<\Phi_0$ (see Fig.~\ref{fig:numericalsimulationsbosons}\textbf{a}) such that in its proximity:
\begin{equation}
\hspace{-0.15cm}
J_c
\sim
\left\{
\begin{array}{ll}
J_{\Phi_1}+C_1\sqrt{\Phi_1-\Phi}+C_2(\Phi_1-\Phi) & (\Phi < \Phi_1) \\
\\
J_{\Phi_1}+C_3 (\Phi-\Phi_1) & (\Phi > \Phi_1)
\end{array} \right.
 \,\, .
\label{eq:currentsingularbehaviorflux}
\end{equation}
The exact values of the coefficients $J_{\Phi_1}$, $C_1$, $C_2$ and $C_3$ can be determined from the numerical simulations. Such singular scaling can be linked to the occurrence of a Lifshitz commensurate-incommensurate (C-IC) transition~\cite{giamarchi2003quantum}, with dynamical exponent $z=2$, and signals the transition from {an incommensurate phase ($\Phi<\Phi_1$) to a commensurate phase ($\Phi>\Phi_1$). The flux $\Phi_1$ identifies the beginning of the helical region.} Similar considerations hold also for a flux value $\Phi_2>\Phi_0$, which identifies the ``end'' of the helical region.

Furthermore, the fractional helical behavior can be distinguished from the other ordinary gapless phases by the value of the \emph{central charge} $c$, which roughly speaking is equal to half the number of gapless modes. Since in the helical case there are only two counter-propagating gapless modes, $c$ is equal to $1$, and $c$ is equal to $2$ in the ordinary gapless liquid because there are four gapless modes. The value of the central charge $c$ of the system can be easily extracted from the entanglement entropy (EE), $S(\ell)$, which we here briefly recall. Let $\hat\rho$ be the density matrix of the system, and $\hat\rho_\ell={\rm Tr}_{L-\ell}[\hat\rho]$ the \emph{reduced} density matrix of a bipartition of the chain, of length $\ell$. The EE is defined by
$
S(\ell)=-{\rm Tr}\left[\hat\rho_\ell\log\left(\hat\rho_\ell\right)\right]$.
For open boundary conditions (OBC), the EE for the ground state (GS) is predicted to scale as~\cite{1742-5468-2004-06-P06002}
\begin{equation}
S(\ell)=s_1+\frac{c}{6}\log\left[\left(\frac{2L}{\pi}\right)\sin\left(\frac{\pi\ell}{L}\right)\right] \,\, ,
\label{eq:entanglemententropycalabresecardy}
\end{equation}
where $s_1$ is a non-universal value. Since $S(\ell)$ can be easily numerically computed, to detect and characterize the $\nu=1/2$ state, we resort on the computation of the chiral current and of the central charge.

\begin{figure}[t]
\centering
\includegraphics[width=8cm]{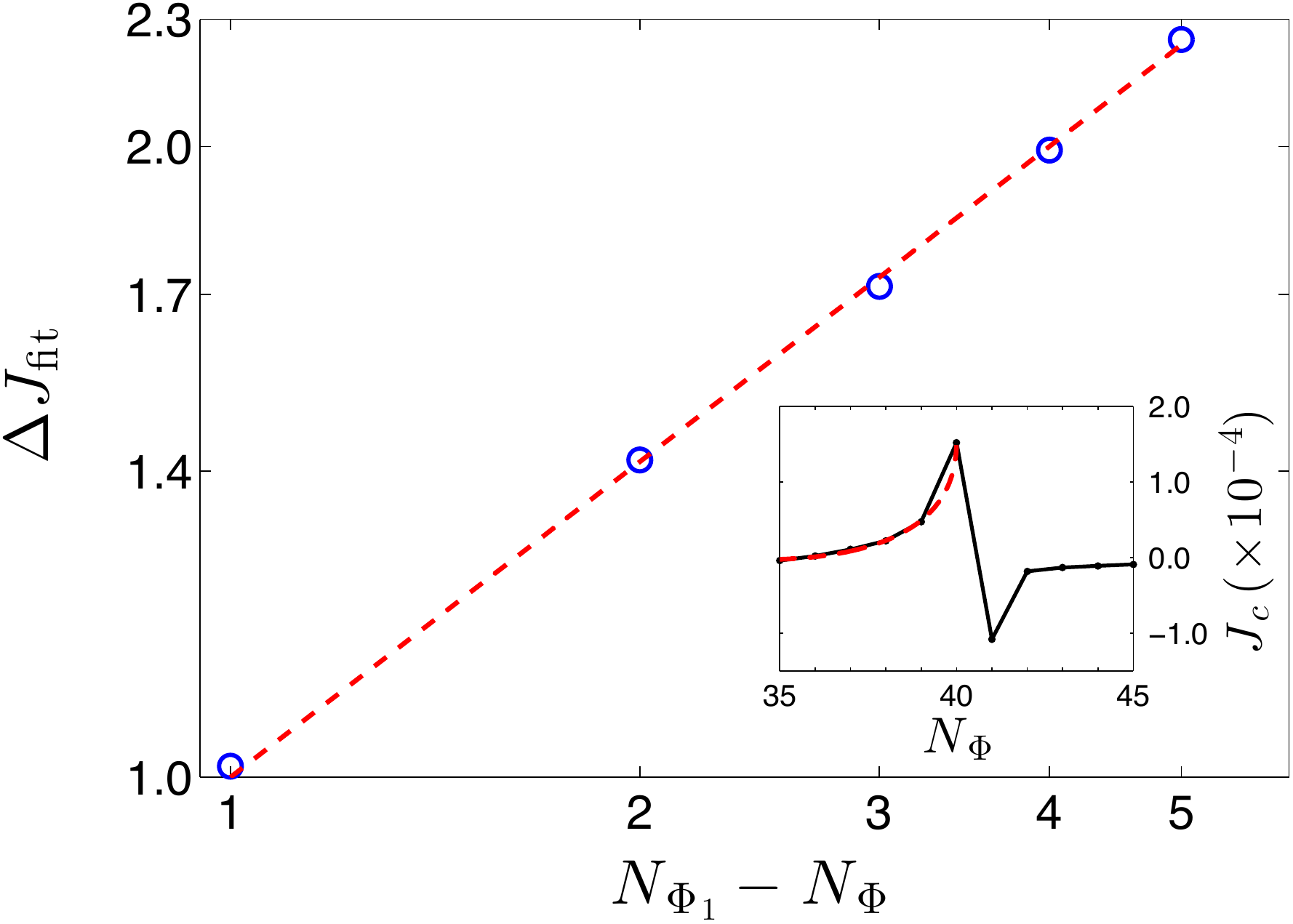}
\caption{(Color online) Fit of the current data presented in Fig.~\ref{fig:numericalsimulationsbosons} for $V_\perp/t=30$, $L=120$, $N=40$, $t_\perp/t=10^{-2}$ (blue dots) using Eq.~\eqref{eq:currentsingularbehaviorflux}, where $\Delta J_{\rm fit}=[J_c(\Phi)-J_\Phi-C_2(\Phi_1-\Phi)]/C_1$ (see text), and $J_{\Phi_1}$ and $\Phi_1$ are taken from the cusp maximum at $N_{\Phi} = 40$. We show the data in log-log scale, and fit for $N_\Phi<N_{\Phi_1}$. In the inset we plot the data in linear scale with the fitted function superimposed.}
\label{fig:databosoniccurrent1}
\end{figure}

\begin{figure*}[t]
\includegraphics[width=5.8cm]{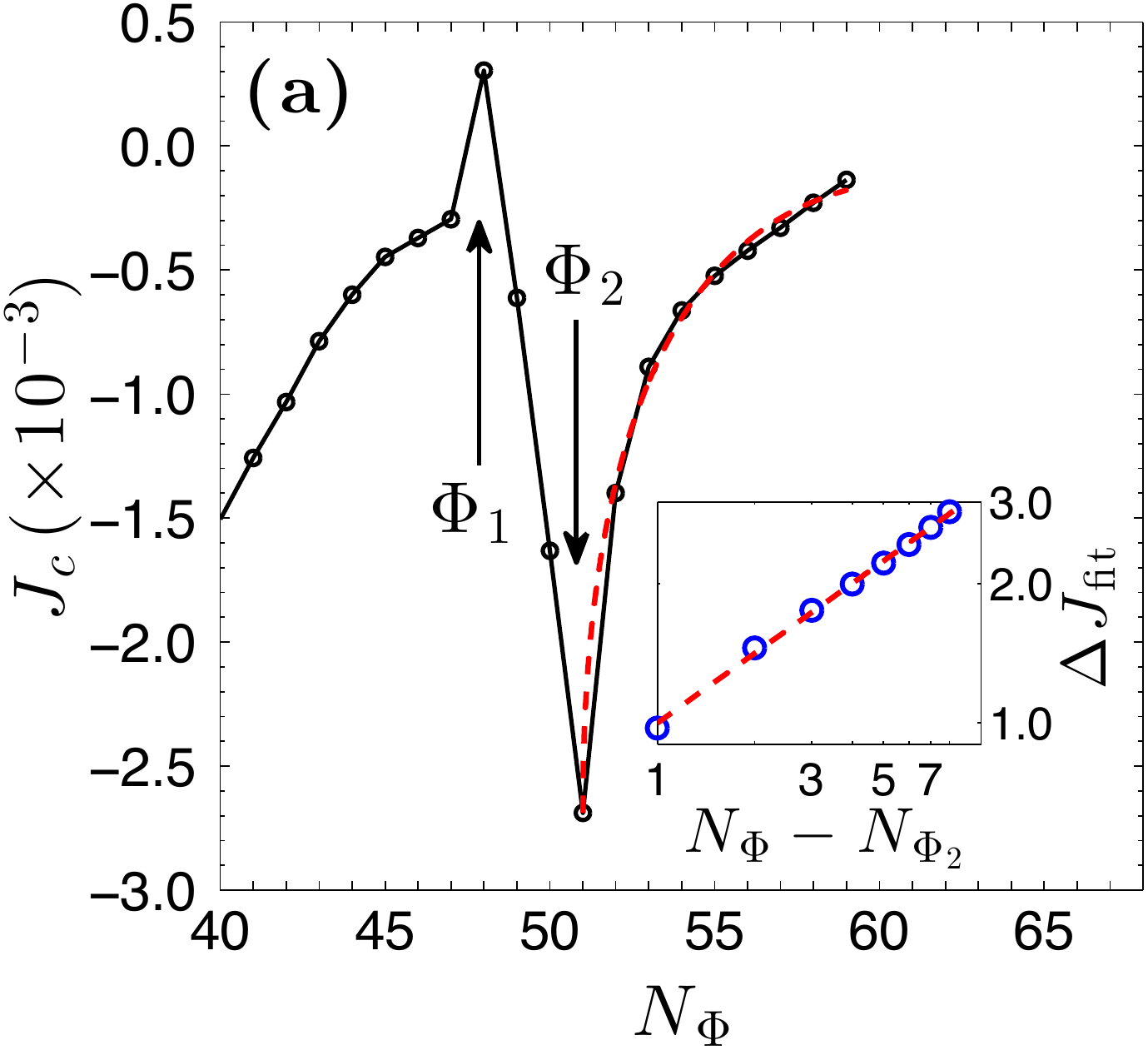}
\includegraphics[width=6.1cm]{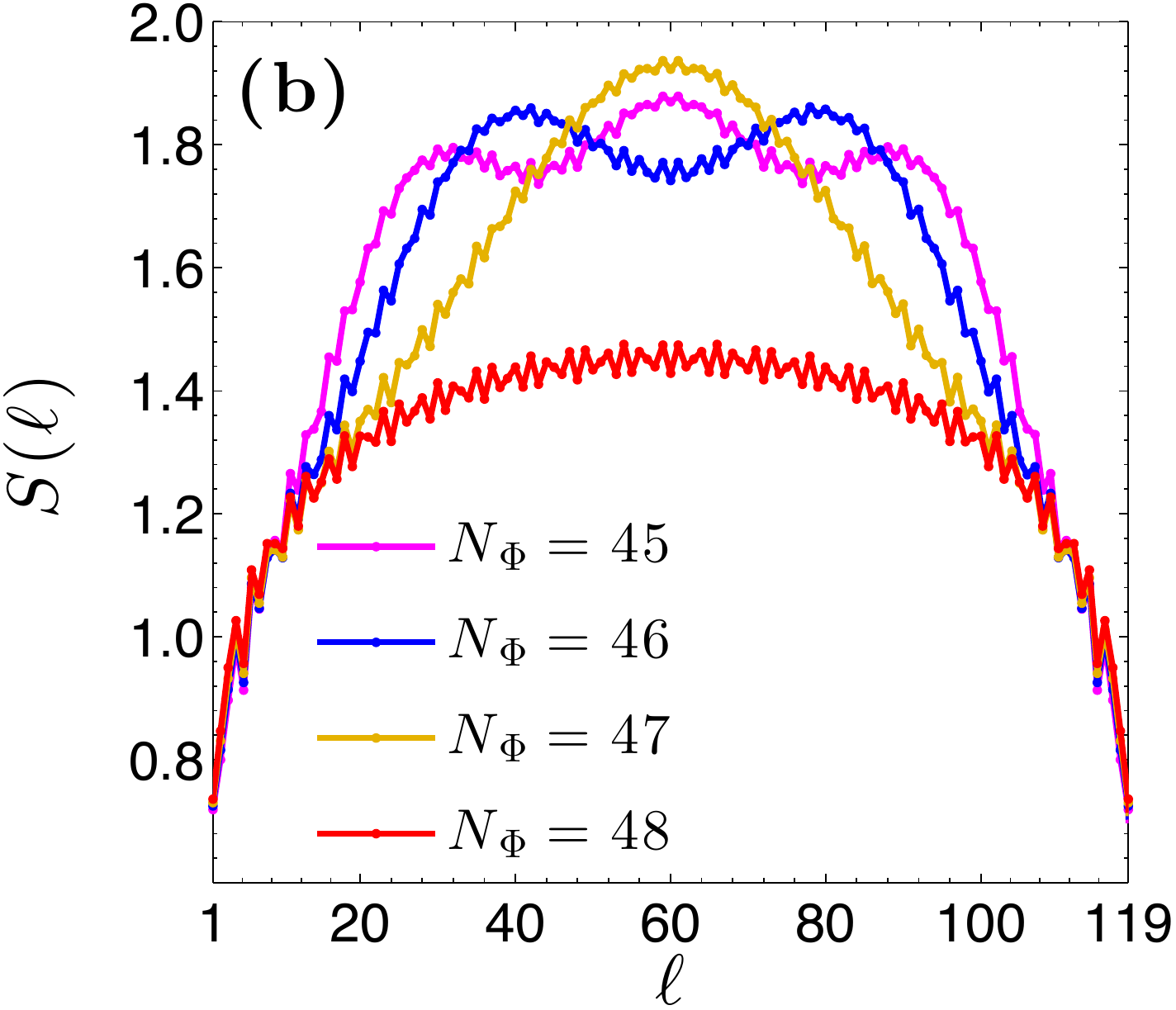}
\includegraphics[width=5.7cm]{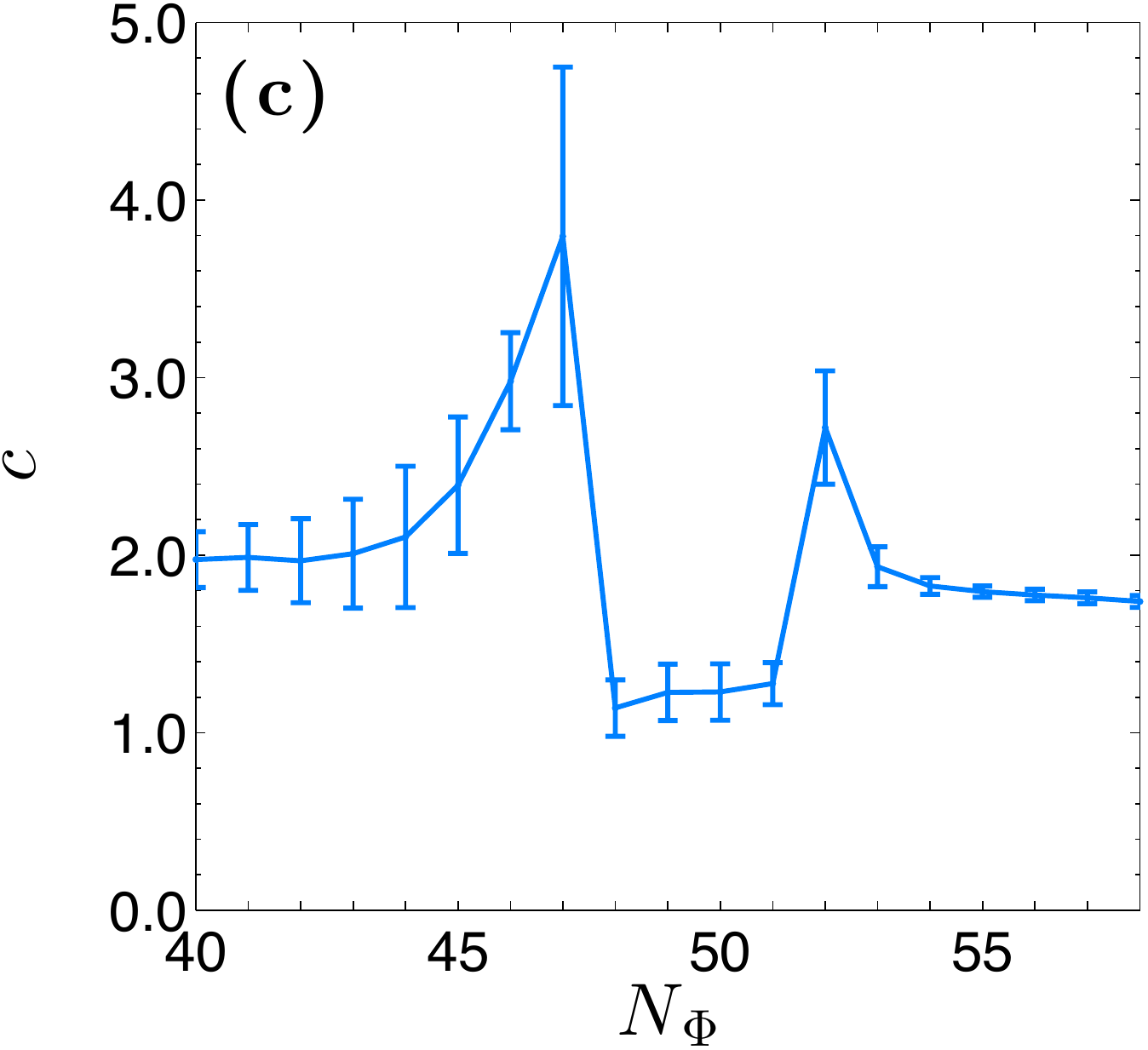}
\caption{(Color online) Numerical results for the $\nu=1/2$ resonance for HCBs, with $L=120$, $N=50$, $t_\perp/t=10^{-1}$, $V_\perp/t=30$, and $D_{\rm max}=400$. \textbf{(a)} Data for $J_c$. We expect the signal of the $\nu=1/2$ Laughlin-like state to occur at $N_\Phi=50$. In the inset, we show the fit with Eq.~\eqref{eq:currentsingularbehaviorflux}, for $N_\Phi>N_{\Phi_2}$ (see text). \textbf{(b)} Entanglement entropy before the C-IC transition, for $N_\Phi=45,46,47,48$. 
\textbf{(c)} Central charge as a function of $N_\Phi$. For $48 \lesssim N_\Phi\lesssim51$, the central charge is equal to one, consistently with the presence of a Laughlin-like state.
}
\label{fig:currentcentralchargeandentanglemententropybosons}
\end{figure*}

\subsection{Numerical results}
\label{sec:numericalresults}
We use an MPS-based algorithm to numerically compute the GS of the system~\cite{Schollwock201196}. We initialize the chain in a random MPS state, and perform an imaginary-time evolution up to time $150\,t^{-1}$ using a fourth-order Trotter expansion, with maximum bond link $D_{\rm max,im}=200$. We then find the GS by means of a variational search in the MPS space, sweeping the chain until convergence is reached (signaled by the constant value approached by the GS energy). We fix the maximum bond link $D_{\rm max}$ during the variational procedure, $N$, $L$, $t_\perp$, $V_\perp$, and vary the magnetic flux, $\Phi$, for each simulation. {To avoid boundary effects}, we discretize the flux as $\Phi=\Phi(N_\Phi)=2\pi N_\Phi/(L+1)$, where $N_{\Phi}=0,\dots,\frac{L}{2}$ is an integer number (for $L$ even). In our simulations, we adopt OBC, and thus compute $J_c$ as the spatial average of the expectation value of spin-resolved current operator $\hat J_{j,m}$, i.e. $J_c=\frac{1}{L-2\Delta L}\sum_{j=\Delta L}^{L-\Delta L}\langle\hat J_{j,+\frac{1}{2}}\rangle$, excluding the $20\%$ of the sites from both chain ends ($\Delta L=0.2\,L$), to avoid boundary effects. The values of $\langle\hat J_{j,m}\rangle$ are given in units of $t$.

The results of simulations for $t_\perp/t=10^{-2}$, $L=120$ and different values {of $N$} are shown in Fig.~\ref{fig:numericalsimulationsbosons}. 
The expected singularities in the current appear precisely at the flux $\Phi =\pi n/\nu$, which scales with the density according to the fractional filling factor $\nu=1/2$, and lead to a peculiar two-cusp pattern: {such a behavior is a clear signature of the occurrence of the Laughlin-like state, which} appears in the small region between the cusps, for $\Phi_1<\Phi<\Phi_2$ (see Fig.~\ref{fig:numericalsimulationsbosons}\textbf{a}).
The narrowness of such region is due to the small value of $t_\perp/t$ which is motivated by the necessity of comparing with bosonization; the finite value of $L$ is responsible for the limited number of numerical points inside it (in this case they are at most two). 
On the other hand, we observe an enhancement of the current as $V_\perp/t$ is increased: this is consistent with general arguments based on bosonization (see Refs.~\cite{PhysRevB.80.165119,PhysRevB.82.045127})  which are reported in Appendix~\ref{sec:bosonizationconventions}.

In Fig.~\ref{fig:databosoniccurrent1}, we present a fit of the current dependence on $\Phi$ with Eq.~\eqref{eq:currentsingularbehaviorflux} for the data series in Fig.~\ref{fig:numericalsimulationsbosons}\textbf{b}, with $V_\perp/t=30$. The fit is performed only for $\Phi<\Phi_1$, fixing $\Phi_1$ and $J_{\Phi_1}$ from the cusp maximum and treating $C_1$ and $C_2$ as fit parameters. In the main panel, to explicitly show the square-root behavior, we plot $\Delta J_{\rm fit}=[J_c-J_{\Phi_1}-C_2(\Phi_1-\Phi)]/C_1$ as a function of $N_{\Phi_1}-N_{\Phi}$. The red curve corresponds to the function $\sqrt{N_{\Phi_1}-N_{\Phi}}$, which well overlaps with our numerical data. In the inset, we show the corresponding current pattern, and the fitting curve with Eq.~\eqref{eq:currentsingularbehaviorflux}. {Our data are consistent with the expected behavior, even if the quality of the fit is limited by the small number of points}. The same analysis can be performed for the other data series in Fig.~\ref{fig:numericalsimulationsbosons}. However, such singular scaling is not evident for $\Phi>\Phi_2$, and this may be due to the small value of $L$ considered (for a critical discussion on the accuracy of the numerical data, see Appendix~\ref{sec:eefit}).

In order to have a larger number of states inside the fractional gap, and thus a clearer characterization of the helical region, we repeated the same simulations as in Fig.~\ref{fig:numericalsimulationsbosons}, with $N=50$, increasing the inter-chain hopping to $t_\perp/t=10^{-1}$. 
Our results are shown in Fig.~\ref{fig:currentcentralchargeandentanglemententropybosons}; in panel~\textbf{(a)}, we demonstrate once more the possibility of fitting the current with the formula in Eq.~\eqref{eq:currentsingularbehaviorflux}.

In this situation we can also perform a numerical analysis of the central charge and of the entanglement properties in the Laughlin-like helical region.
We fit the central charge via Eq.~\eqref{eq:entanglemententropycalabresecardy}; since the obtained EE has an oscillating behavior (Fig.~\ref{fig:currentcentralchargeandentanglemententropybosons}\textbf{b}), we fix $s_1$ and identify the extremal values $s_{1,\rm min}$ and $s_{1,\rm max}$ for which the fit intercepts the oscillating behavior of $S(\ell)$. Accordingly, we obtain two related values $c_{\rm min}$ and $c_{\rm max}$. The mean value is then plotted, and its uncertainty is roughly estimated via $(c_{\rm max}-c_{\rm min})/2$ (see Appendix~\ref{sec:eefit}). Note that the behavior of the current is paralleled by the different values of the central charge in the three regions $\Phi<\Phi_1$, $\Phi_1<\Phi<\Phi_2$ and $\Phi>\Phi_2$. 
In the vicinity of $\Phi_1$, the EE develops pronounced oscillations, in the presence of OBC, which make the fit with Eq.~\eqref{eq:entanglemententropycalabresecardy} fail. The observed behavior of the EE across the transition has been also detected in other models~\cite{PhysRevB.94.115112,PhysRevB.87.115132,PhysRevB.92.035154}, and ascribed to the occurrence of the Lifshitz transition.

The fact that we do not exactly fit $c=1$ or $c=2$ in the helical and non-helical region respectively, and the large uncertainties on the values of $c$, (see Fig.~\ref{fig:currentcentralchargeandentanglemententropybosons}\textbf{c}) can be due to finite-size and boundary effects, as well as to numerical errors during the variational minimization. In particular, we impute this disagreement to the finite values of the maximum bond link, $D_{\rm max}=400$, that we use to describe the MPS state at the end of the variational procedure. Because of the large amount of entanglement, to correctly describe the MPS state during and at the end of the variational procedure, we see that we would need $D_{\rm max}>400$, which can not be achieved with our computational resources.

With $D_{\rm max}=400$, we see that the smallest singular eigenvalue at the center of the system ($j=L/2$), at the end of the variational procedure, is of the order of $10^{-6}$, which identifies our truncation error. The fact that we need to truncate the MPS state affects the computation of the EE~\cite{Schollwock201196}, but has a less drastic effect on the computation of the current (see Appendix~\ref{sec:eefit}). In the light of these results, we judge reliable our numerical data for the chiral current.

\begin{figure*}
\centering
\includegraphics[width=18cm]{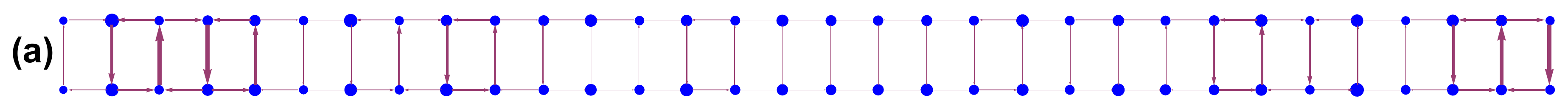}
\includegraphics[width=18cm]{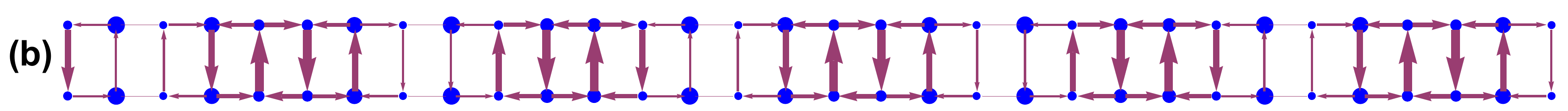}
\includegraphics[width=18cm]{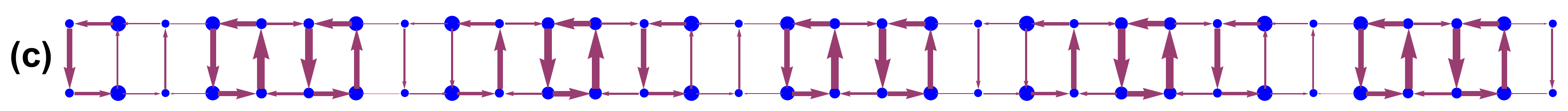}
\includegraphics[width=18cm]{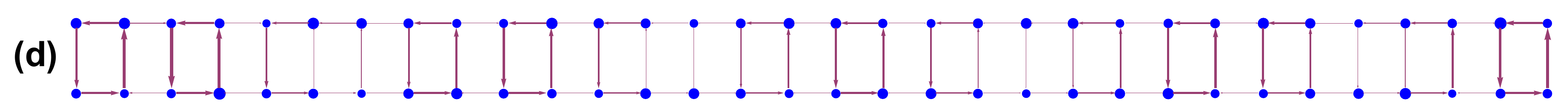}
\caption{{(Color online) Current configuration along the ladder for the data in Fig.~\ref{fig:currentcentralchargeandentanglemententropybosons}\textbf{a}. \textbf{(a)} Data for $N_\Phi=47$ (just before the beginning of the helical region), \textbf{(b)} for $N_\Phi=48$ (just after the beginning of the helical region), \textbf{(c)} for $N_\Phi=51$ (just before the end of the helical region), and \textbf{(d)} for $N_\Phi=52$ (just after the end of the helical region). Blue dots on the sites of the ladder represent the local densities $\langle\hat n_{j,m}\rangle$.}}
\label{fig:currentconfigurationalongtheladder}
\end{figure*}

We also notice that the {chiral current} (Fig.~\ref{fig:currentcentralchargeandentanglemententropybosons}\textbf{a}) does not vanish at $N_\Phi\simeq50$, and does not obey the square-root scaling for $N_\Phi<47$. Both these features, can be attributed to the presence of an additional background signal which influences that of the $\nu=1/2$ Laughlin-like state. A deeper characterization of the nature of this signal is beyond the aim of this manuscript. We just mention that the chiral current in the bosonic ladder has been recently studied in Refs.~\cite{PhysRevB.91.140406,PhysRevLett.115.190402,PhysRevA.94.063628}.

{We show in Fig.~\ref{fig:currentconfigurationalongtheladder} the spatial configuration of the current for the data in Fig.~\ref{fig:currentcentralchargeandentanglemententropybosons}, for four different values of $N_\Phi$, focusing on 15 rungs centered around $L/2$. The currents along the legs is computed by using the operator in Eq.~\eqref{eq:spinresolvedcurrentoperator}, whereas the inter-leg (transverse) current is computed via the transverse current operator:
\begin{equation}
\hat J_{\perp,j}=-it_\perp\left(\hat b^\dag_{j,-\frac{1}{2}}\hat b_{j,+\frac{1}{2}}\,e^{-i\Phi j}-{\rm H.c.}\right) \,\, .
\end{equation}
The data are shown for {\textbf{(a)}} $N_\Phi=47$ (just before the beginning of the helical region), \textbf{(b)} for $N_\Phi=48$ (just after the beginning of the helical region), \textbf{(c)} for $N_\Phi=51$ (just before the end of the helical region), and \textbf{(d)} for $N_\Phi=52$ (just after the end of the helical region). From our data, we see that there is a sudden change of the current pattern when the system enters the Laughlin-like phase, where an ordered current configuration is numerically observed. Such an order is also observed in the density modulation $\langle\hat n_{j,m}\rangle$, represented by blue dots.}

\begin{figure}[t]
\centering
\includegraphics[width=8cm]{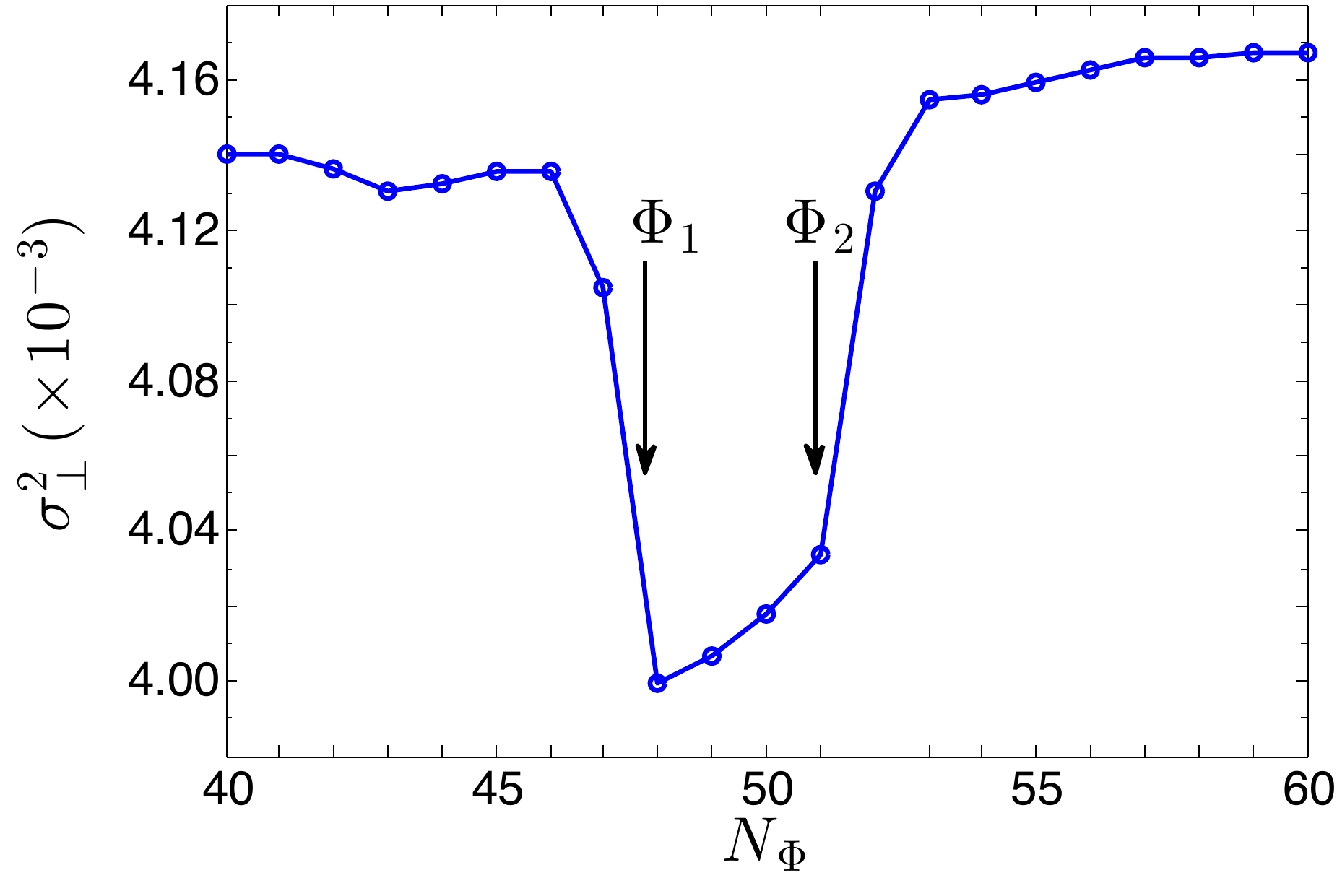}
\caption{{(Color online) Transverse current fluctuations as a function of $N_\Phi$ for the data in Fig.~\ref{fig:currentcentralchargeandentanglemententropybosons}. We see that the transverse current fluctuations are minimized inside the Laughlin-like region, in agreement with bosonization predictions.}}
\label{fig:transversecurrentfluctuations}
\end{figure}

{We can try to understand the emergence of an ordered current pattern by using bosonization (see Appendix~\ref{sec:bosonizationconventions}). Within such framework, we obtain that $\langle\hat J_\perp(x)\rangle=0$, which is consistent with the data in Fig.~\ref{fig:currentconfigurationalongtheladder}, if such expectation value is intended in a coarse-grained sense as an average over several sites.}

{Furthermore, the transverse current fluctuations should be suppressed in the Laughlin-like state. We then define $\sigma_{\perp,j}^2=\langle\hat J_{\perp,j}^2\rangle-\langle\hat J_{\perp,j}\rangle^2$, and we expect $\sigma^2_{\perp,j}$ to be minimized within the Laughlin-like region. We show the result in Fig.~\ref{fig:transversecurrentfluctuations}, where we plot $\sigma^2_\perp=\frac{1}{L-2\Delta L}\sum_{j=\Delta L}^{L-\Delta L}\sigma^2_{\perp,j}$ for the data in Fig.~\ref{fig:currentcentralchargeandentanglemententropybosons}. As before, spatial averages are computed by discarding the $20\%$ of the chain from both ends to avoid boundary effects. From our data, the transition to the Laughlin-like state is signaled by a suppression of $\sigma^2_\perp$ when entering the helical phase.}

Despite the numerical complexity of the problem, our analysis shows clear signatures of the $\nu=1/2$ Laughlin-like state via the observation of the chiral current and entanglement properties.

\begin{figure*}[t]
\includegraphics[width=5.8cm]{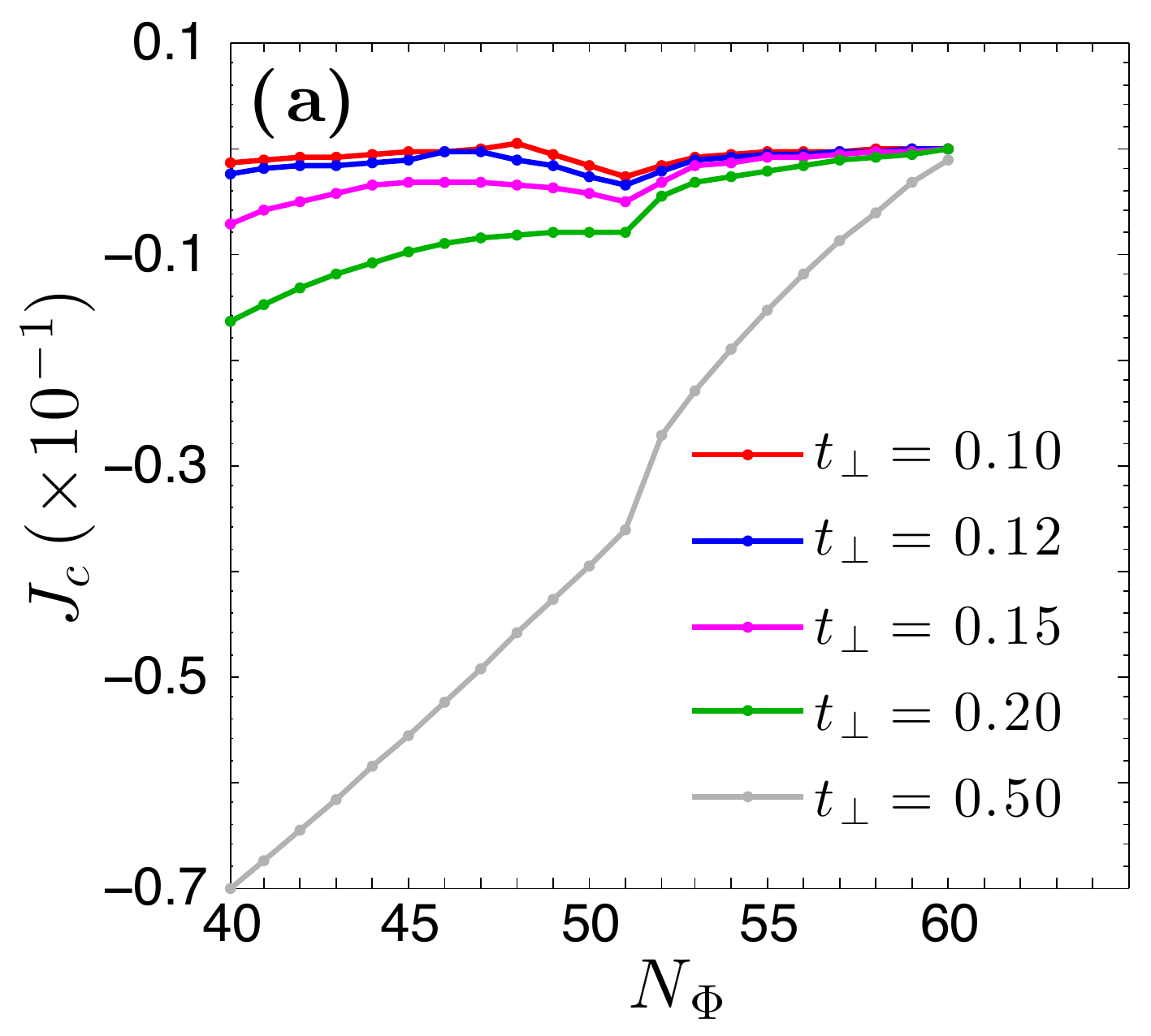}
\includegraphics[width=5.8cm]{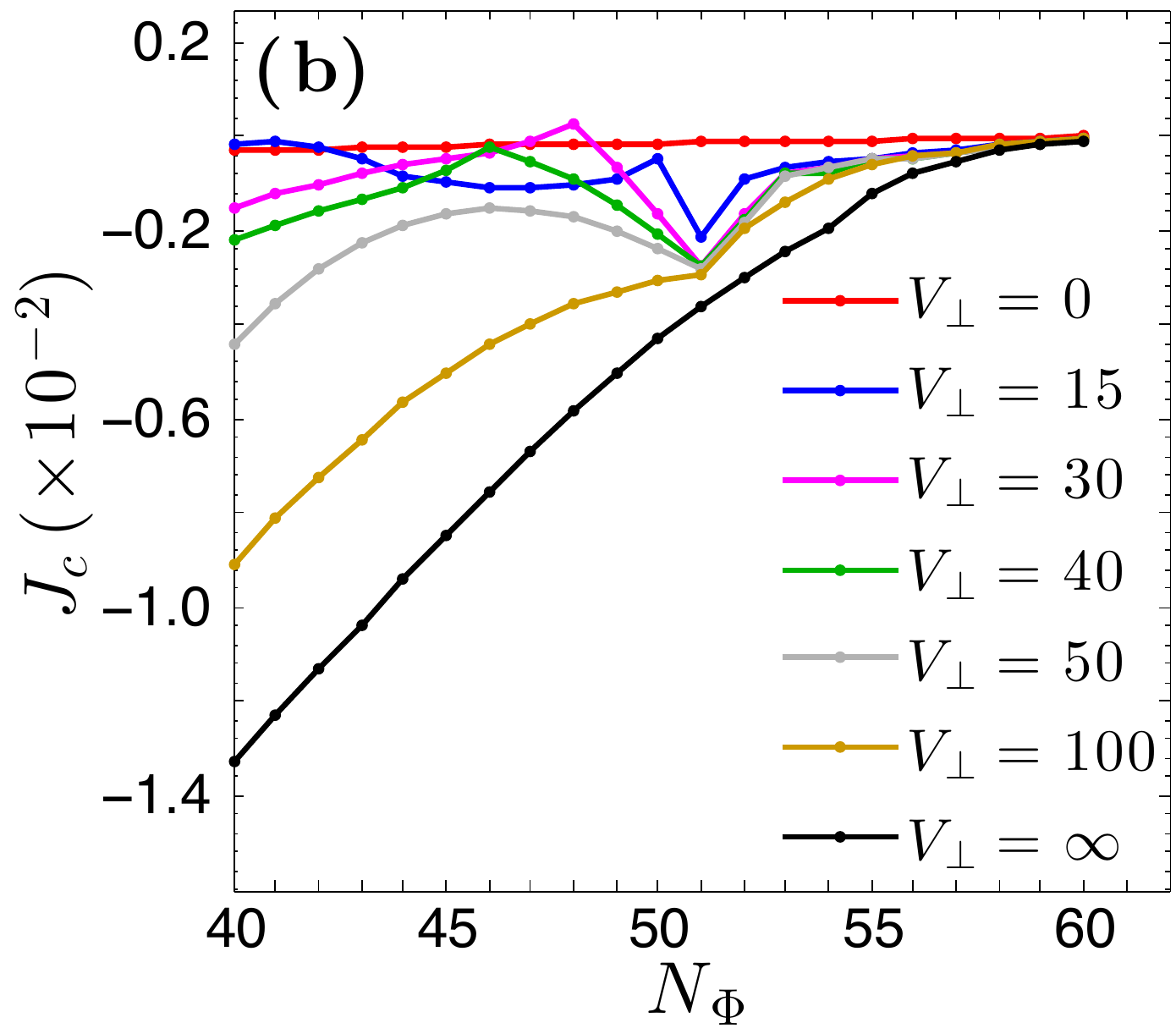}
\includegraphics[width=6.1cm]{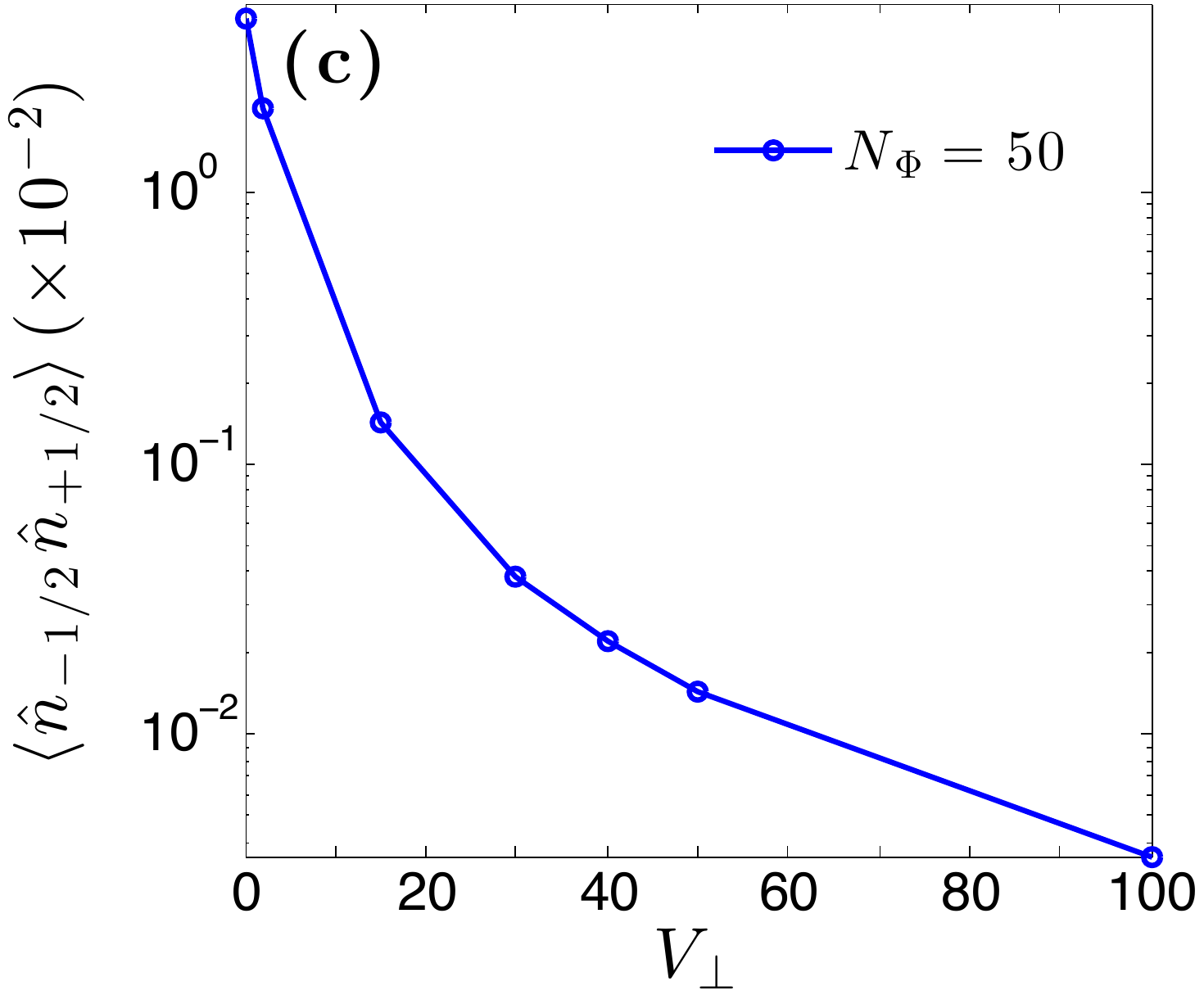}
\caption{{(Color online) Numerical results for the $\nu=1/2$ resonance for HCBs, with $L=120$, $N=50$, and $D_{\rm max}=200$. \textbf{(a)} Data for $J_c$ at $V_\perp/t=30$ and for different values of $t_\perp/t$. \textbf{(b)} $J_c$ at $t_\perp/t=0.1$ for different values of $V_\perp/t$. \textbf{(c)} Double-occupation probability $\langle\hat n_{-\frac{1}{2}}\hat n_{+\frac{1}{2}}\rangle$ as a function of $V_\perp/t$ for the data in panel \textbf{(b)} with $N_\Phi=50$.}}
\label{fig:currenttperpandvperpscan}
\end{figure*}

\subsection{{Interleg couplings}}
\label{sec:interwirecouplings}
{The analysis presented in the previous sections is valid if we are in the regime of parameters in which bosonization is valid, i.e., $t_\perp/t\ll1$, and $V_\perp/t$ large enough to ensure the occurrence of the Laughlin-like state. Here, we extend our results exploring the whole range of values for $t_\perp/t$ and $V_\perp/t$, from small to large ones. In Fig.~\ref{fig:currenttperpandvperpscan}\textbf{a}, we show the data for $J_c$ for $L=120$, $N=50$, $V_\perp/t=30$ and $D_{\rm max}=200$, for different values of $t_\perp/t$. We see that, for this choice of parameters, the double-cusp pattern is lost for $t_\perp/t\gtrsim0.1$. For larger $t_\perp/t$, a discontinuity around $N_\Phi=50$ is observed, which might still indicate the presence of the Laughlin-like state. The chiral current for different $V_\perp/t$, at $t_\perp/t=0.1$, is shown in Fig.~\ref{fig:currenttperpandvperpscan}\textbf{b}. The double-cusp pattern is visible up to $V_\perp/t=30$, and it starts to disappear for larger values of $V_\perp/t$. The data at $V_\perp/t=\infty$ are taken by implementing the hard-core limit as a constraint by reducing the dimension of the local Hilbert space. From these simulations, we notice that increasing $t_\perp/t$ and/or $V_\perp/t$ over some threshold values destroys the double-cusp pattern of $J_c$. Thus, in order to observe it, we have to keep $t_\perp/t\lesssim0.1$, whereas $V_\perp/t$ has to be sufficiently large, but finite.}

{Finally, in order to check that the disappearance of the cusp for large $V_\perp/t$ is not a numerical artifact, we computed the double-occupation probability (with HCBs there can not be more than two particles per rung). The expectation value $\langle\hat n_{-\frac{1}{2}}\hat n_{+\frac{1}{2}}\rangle=L^{-1}\sum_j\langle\hat n_{j,-\frac{1}{2}}\hat n_{j,+\frac{1}{2}}\rangle$ is shown in Fig.~\ref{fig:currenttperpandvperpscan}\textbf{c}; it refers to the data in Fig.~\ref{fig:currenttperpandvperpscan}\textbf{b} for $N_\Phi=50$ (all curves with different $N_\Phi$ overlap, not shown). The double-occupation probability is always decreasing and correctly goes to zero when approaching the large-$V_\perp$ limit.}

\section{Laughlin-like states in fermionic ladders}
\label{sec:laughlinstatesinfermionicmodels}

In this section, we discuss Laughlin-like states in fermionic synthetic ladders. The experimental realization of such ladders has been reported in Refs.~\cite{science1510, PhysRevLett.117.220401}.
Bosonization predicts that fermionic Laughlin-like states appear when the filling factor, defined as in the bosonic case, is $\nu=1/p$, and $p$ is an \emph{odd} positive integer~\cite{PhysRevLett.88.036401,PhysRevB.89.085101,PhysRevB.92.115446}, provided that the range of interactions is sufficiently long. As we did in the bosonic case, here we focus on the characterization of the most easily accessible Laughlin-like state, which is the $\nu=1/3$ state.

\subsection{Model}

We model the synthetic fermionic ladder with
the Hamiltonian $\hat H=\hat H_0+\hat H_\perp + \hat H_{\rm int}$, where
\begin{subequations}
\label{eq:hamiltonianfermionicsystem}
\begin{align}
\hat H_0 &=-t\sum_j\sum_{m=\pm1/2}\left(\hat a^\dag_{j,m}\hat a_{j+1,m}+\mathrm{H.c.}\right) \, ; \label{eq:hamiltonianlatticenogaugeshift1}\\
\hat H_\perp &=
+t_\perp\sum_{j}\left(e^{-i\Phi j}\,\hat a^\dag_{j,-\frac{1}{2}}\hat a_{j,+\frac{1}{2}}+\mathrm{H.c.}\right) \, ; \label{eq:hamiltonianlatticenogaugeshift}\\
\hat H_{\rm int} &=U\sum_j\hat n_{j,+\frac{1}{2}}\hat n_{j,-\frac{1}{2}}+V\sum_j\hat n_j\hat n_{j+1} \, .
\label{eq:hamiltonianlatticenogaugeshiftinteracting}
\end{align}
\end{subequations}
Here, $\hat a_{j,m}$ ($\hat a^\dag_{j,m}$) is the fermionic annihilation (creation) operator on site $j$, satisfying the usual anti-commutation relations, $\{\hat a_{j,m},\hat a^\dag_{j',m'}\}=\delta_{j,j'}\,\delta_{m,m'}$; $\hat n_{j,m} = \hat a_{j,m}^\dagger \hat a_{j,m}$ and $\hat n_j=\sum_{m=\pm1/2}\hat n_{j,m}$. The hopping parameters $t$ and $t_\perp$ are as in the bosonic case, and $U$ and $V$ denote the on-site and nearest-neighbor (NN) interaction strength, respectively. Again, we use $t$ as reference energy scale. By computing the chiral current as a function of $\Phi$, we expect to observe both the signal of the $\nu=1$ helical state, at $\Phi=\pi n$ ($N_\Phi\simeq N/2$), and the one of the $\nu=1/3$ state, at $\Phi=3\pi n$ ($N_\Phi\simeq3N/2$). 

\subsection{Numerical results}

Before commenting on the occurrence of the $\nu=1/3$ Laughlin-like state, we need to discuss in some detail the effect of interactions on the $\nu = 1$ resonance, which appears also in simple free-fermion models.
As predicted by bosonization (see Appendix~\ref{sec:bosonizationconventions}), and as also discussed in Refs.~\cite{PhysRevB.80.165119,PhysRevB.82.045127}, repulsive interactions in Eq.~\eqref{eq:hamiltonianfermionicsystem} enhance the chiral-current signal of the non-interacting $\nu=1$ helical state, thus yielding an effective single-particle spectrum with a renormalized value of $t_\perp$ which is enhanced by interactions. 

\begin{figure}[t]
\centering
\includegraphics[width=8.2cm]{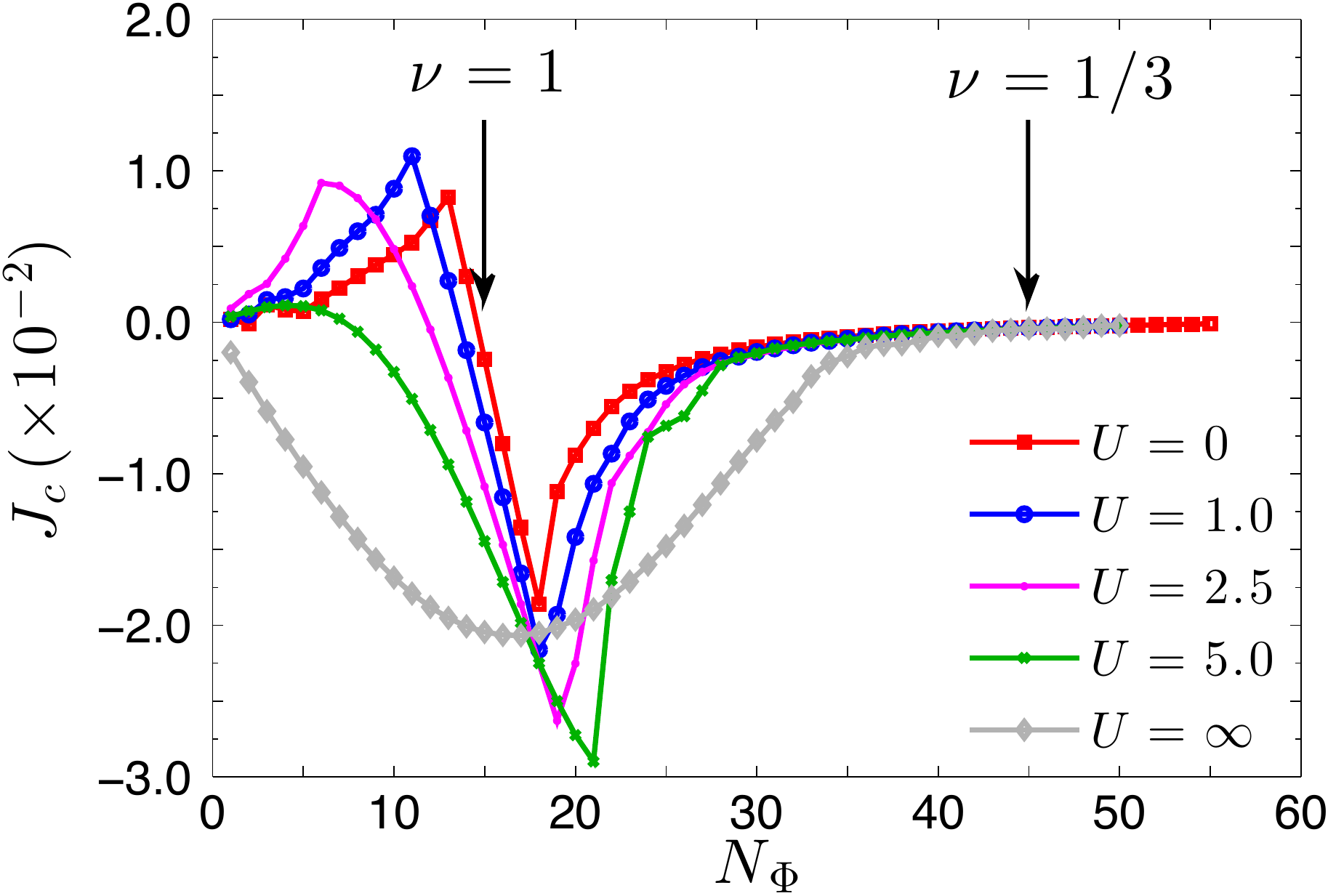}
\caption{(Color online) Chiral current for interacting fermions for different values of $U/t$, and $V/t=0$ in Eq.~\eqref{eq:hamiltonianfermionicsystem}. Here we use $L=120$, $N=30$, $t_\perp/t=10^{-1}$ and vary $U$ as in the legend. The series with $U=\infty$ has been taken by implementing the hard-core limit as a reduction of the dimension of the local Hilbert space, on each site. The arrows indicate the expected values of $N_\Phi$ at which the $\nu=1$ and $\nu=1/3$ resonances occur (the latter expected only if $V\neq0$), which are $N_\Phi\simeq15$ and $N_\Phi\simeq45$ respectively.}
\label{fig:chiralcurrentgaprenormalization}
\end{figure}

In order to observe the effect of interactions on the current signal of the $\nu=1$ helical state, we simulate Hamiltonian in Eq.~\eqref{eq:hamiltonianfermionicsystem} with the MPS-based algorithm described in Sec.~\ref{sec:numericalresults} in the limit of on-site interaction only ($V=0$); results are shown in Fig.~\ref{fig:chiralcurrentgaprenormalization}. We use $L=120$, $N=30$, $t_\perp/t=10^{-1}$ and vary $U/t$ as in the legend ($U=\infty$ is obtained by reducing the dimension of the local Hilbert space). With these parameters, the signal of the $\nu=1$ state arises at $N_\Phi\simeq15$, as expected. We see that the size of helical region increases with $U/t$. When $U/t\gtrsim2.5$,  the helical regime begins to be less discernable, and clearly disappears for large values of $U/t$. The same behavior is observed if we simulate Hamiltonian in Eq.~\eqref{eq:hamiltonianfermionicsystem} with $U=V=0$, and gradually increase $t_\perp/t$ (not shown). We thus conclude that the behavior of $J_c$ we observe in Fig.~\ref{fig:chiralcurrentgaprenormalization} reflects the renormalization of the non-interacting gap, complementing previous observation on the momentum distribution function~\cite{1367-2630-18-3-035010}.

When NN interactions are turned on, the signal of the $\nu=1/3$ Laughlin-like state should occur at $N_\Phi\simeq 45$ for the parameters used in Fig.~\ref{fig:chiralcurrentgaprenormalization}. However, we have not succeeded in providing unambiguous evidence of the occurrence of the $\nu=1/3$ Laughlin-like state from the direct simulation of Hamiltonian in Eq.~\eqref{eq:hamiltonianfermionicsystem}. We ascribe this difficulty to the enhancement of the $\nu=1$ signal {due to interactions, as well as to the intrinsic numerical difficulty of the problem}.

\subsection{Exact mapping}
\label{sec:exactmapping}

In order to explicitly observe the $\nu=1/3$ Laughlin-like state {in a fermionic ladder, we consider another model. The results which we are going to present do not mathematically imply that the $\nu=1/3$ Laughlin-like state also appears in model~\eqref{eq:hamiltonianfermionicsystem}, but constitute an evidence that bosonization predictions of Laughlin-like states in fermionic ladders have a range of validity.}

The Hamiltonian is ${\hat H_{\rm ex} = \hat H_0+\hat H_\perp+\hat H_{\mathrm{ex-int}}}$, where $\hat H_0$ and $\hat H_\perp$ are given by Eq.~\eqref{eq:hamiltonianlatticenogaugeshift1} and Eq.~\eqref{eq:hamiltonianlatticenogaugeshift} respectively, and
\begin{equation}
\label{intmodel}
\hat H_{\mathrm{ex-int}} = \sum_j\sum_{r\geq0} V(r)\,\hat n_j\hat n_{j+r} \,\, .
\end{equation}
The Hamiltonian $\hat H_{\rm ex}$ is equal to that in Eq.~\eqref{eq:hamiltonianfermionicsystem} apart from the form of the interaction term. In the following, we specialize to an interaction potential that vanishes beyond the interaction range $\xi$,
\begin{equation}
\label{interaction}
V(r) =
\begin{cases} U   & \mbox{for } r \le \xi  \\ 0  & \mbox{for } r > \xi \end{cases} \,\, ,
\end{equation}
and we consider $U\gg t$ (hard-core limit). These kind of Hubbard-like models~\cite{PhysRevLett.111.165302} are of direct relevance to cold atom experiments where GS atoms are weakly admixed to highly-excited Rydberg states~\cite{PhysRevLett.104.195302,PhysRevLett.105.160404}. Such mixing induces effective interactions between atoms of the form in Eq.~\eqref{interaction}, which have been recently probed via spectroscopy in Refs.~\cite{NatPhys1271,NatPhys121095}.

The solution of this model has been provided in Ref.~\cite{PhysRevB.92.115446} and is also reported for convenience in Appendix~\ref{sec:solutiontotheexactmodel} {(see also Refs.~\cite{PhysRevB.84.085434,arXiv:1607.07842})}. The key idea is to remap the original model to a similar one (primed) with $\xi' = 0$ and different length $L'=L-(N-1)\xi$; the chiral current of the remapped system is related to that of the original system via: 
\begin{equation}
\label{eq:currentrelation2}
J_c\{\nu\}=\frac{L'}{L} J'_c\{\nu'\} \,\, .
\end{equation}
In the thermodynamic limit, we can exactly map a fractional $\nu=1/p$ phase, where $p$ is an odd (even) positive integer for fermions (bosons), onto a $\nu'=1$ phase, or \textit{vice versa}. We can thus obtain the chiral current for the fractional phase by making a numerical simulation for the integer phase $\nu' = 1$, which is computationally much more efficient.

The mapping we presented is exact only at $U/t\rightarrow\infty$. However, this infinite interaction point is somewhat pathological as it contains extra degeneracies. Configurations where subsequent particles along the chain are placed in either $m=-1/2$ or $m=1/2$ legs become degenerate at $U/t\rightarrow\infty$.
In order to avoid this problem, we will turn to an approximation by computing the current for a model at $\nu'=1$ at large but finite interaction $U/t$. Then, Eq.~(\ref{eq:currentrelation2}) becomes approximated. We will check the validity of this approximation by computing the probability of finding two particles in the same site in the numerically-simulated problem; when this probability is small, then the hard-core limit is effectively reached.

\begin{figure}[t]
\centering
\includegraphics[width=8.4cm]{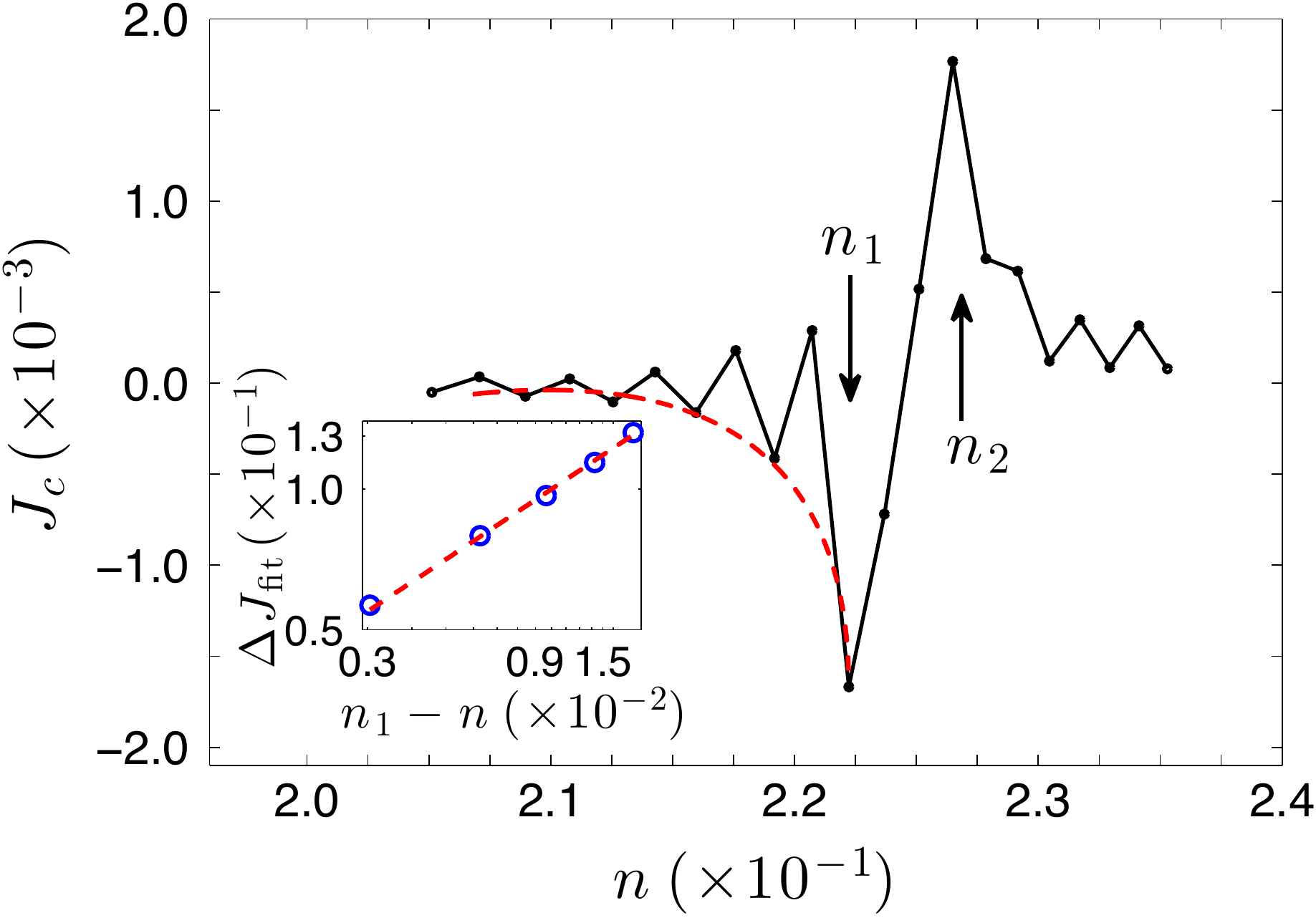}
\caption{(Color online) Chiral current for the $\nu=1/3$ Laughlin-like state for the model $\hat H_{\rm ex}$. The actual system is  at $\nu=1/3$ with $\xi=3$, $L=225$, $\Phi \simeq 0.658 \pi$. The data are obtained through a simulation of a $\nu'=1$ phase using $L'=78$, $N'_{\Phi}=26$ ($\Phi\simeq0.658\,\pi$), $t_\perp/t=10^{-2}$ and $U/t=10$. In the inset, we highlight the behavior described in Eq.~\eqref{eq:currentsingularbehaviordensityreal}.
}
\label{fig:remappedcurrent}
\end{figure}

The numerical results are presented in Fig.~\ref{fig:remappedcurrent}. We compute $J_c'$, simulating the $\nu'=1$ phase of the remapped system using $L'=78$, $N'_\Phi=26$ ($\Phi \simeq 0.658\,\pi$), $U/t=10$, $\xi'=0$, and $t_\perp/t=10^{-2}$. We vary the density $n'=N/L'$ around the expected resonance value, i.e. $n'=\Phi/\pi\simeq0.658$ by sweeping $N$ from $N=40$ to $N=60$ (all quantities listed in this paragraph without a prime are invariant under the mapping, see Appendix~\ref{sec:solutiontotheexactmodel}). Finally, these data are reinterpreted in terms of the original model at $\nu=1/3$ and plotted in the figure.

As previously stated, the fact that we use a finite value of $U/t$ forces us to introduce a prescription to test whether we are in the hard-core limit or not. We then define the Fock basis of the local Hilbert space on site $j$ of the ladder as $\{|sr\rangle_j\equiv|s\rangle_{j,+\frac{1}{2}}\otimes|r\rangle_{j,-\frac{1}{2}}\}$, where $s,r=0,1$. Let $\mathcal{\hat P}_{s,r;j}\coloneqq|sr\rangle_j\langle sr|$ be the projector over the state $|sr\rangle_j$. The total density operator on site $j$ can be written as $\hat n'_j=\sum_{s,r=0,1}(s+r)\mathcal{\hat P}_{s,r;j}$. In the limit $U\gg t$, the double-occupation probability is largely suppressed, i.e. $\langle\hat n'_j\rangle\gg\langle\mathcal{\hat P}_{1,1;j}\rangle$, for all $j$, where the dependence on $j$ arises from the choice of OBC. When this condition is fulfilled, we approach the hard-core limit. Regarding Fig.~\ref{fig:remappedcurrent}, for each plotted value, we checked that the hard-core condition is fulfilled: in our simulations $0.01\lesssim\langle\mathcal{\hat P}_{1,1;j}\rangle/\langle\hat n'_j\rangle\lesssim0.02$, for all $j$.
 
The singular scaling of the current discussed in previous sections is apparent also in Fig.~\ref{fig:remappedcurrent}; contrary to the previous cases, here the density is varied and the flux is fixed, i.e.
\begin{equation}
J_c \sim J_{n_1}+B_1\sqrt{n_1-n}+B_2(n_1-n) \,\, , \quad (n < n_1)
\label{eq:currentsingularbehaviordensityreal}
\end{equation}
for some coefficients $J_{n_1}$, $B_1$, and $B_2$. Here, $n_1$ is the lowest value of $n$ at which the system enters the Laughlin-like state; a similar discussion holds for density larger than a value $n_2>n_1$. Our data confirm the singular scaling in Eq.~\eqref{eq:currentsingularbehaviordensityreal}: 
since $J_c$ displays an oscillating pattern between even and odd values of $N$, we fit only the even values of $N$,
see Fig.~\ref{fig:remappedcurrent} (inset).

Summarizing, we have described {a model characterized by finite-range interactions} which gives {access in} a simple {way} to Laughlin-like states {in fermionic ladders}. {Its solution is based on a mapping that} allowed us to compute the current in a fractional state using the results of an integer quantum-Hall-like state (note that it can also be extended to HCBs). We find it a powerful way to {verify the existence of} Laughlin-like states, since the direct observation for the original model in Eq.~\eqref{eq:hamiltonianfermionicsystem} is much more challenging with the available numerical techniques.

\section{Free-fermion analogy}
\label{sec:IQHE}

In this and in the following section, we motivate the choice of the chiral current and of the central charge as identifiers of the fractional Laughlin-like state. Here, we focus on the fact that the observed features at $\nu=1/2$ and $\nu=1/3$ have an analogy with several features of a free-fermion model at $\nu = 1$.

\subsection{Model}
We consider the Hamiltonian in Eq.~\eqref{eq:hamiltonianfermionicsystem} with $U=V=0$. In order to make it translationally invariant, we perform the unitary transformation $\hat a_{j,m}\rightarrow~e^{i\Phi jm}\,\hat a_{j,m}$ and then move to momentum space using $\hat a_{k,m}=L^{-1/2}\sum_je^{ikj}\,\hat a_{j,m}$, so that it becomes
\begin{equation}
\hat H=\sum_k\left(
\begin{array}{cc}
\hat a^\dag_{k,+\frac{1}{2}} & \hat a^\dag_{k,-\frac{1}{2}}
\end{array}
\right)
\left(
\begin{array}{cc}
\varepsilon_{+\frac{1}{2}}(k) & t_\perp\\
t_\perp & \varepsilon_{-\frac{1}{2}}(k)
\end{array}
\right)
\left(
\begin{array}{c}
\hat a_{k,+\frac{1}{2}}\\
\hat a_{k,-\frac{1}{2}}
\end{array}
\right)
\label{eq:hamiltonianinmomentumspacematrix}
\end{equation}
where $\varepsilon_m(k)=-2t\cos(k-m\Phi)$. 
Its diagonalization yields the two energy bands $E_\pm(k)=-2t\cos(k)\cos(\Phi/2)\pm\sqrt{4t^2\sin^2(k)\sin^2(\Phi/2)+t_\perp^2}$.

\begin{figure}[t]
\centering
\includegraphics[width=8cm]{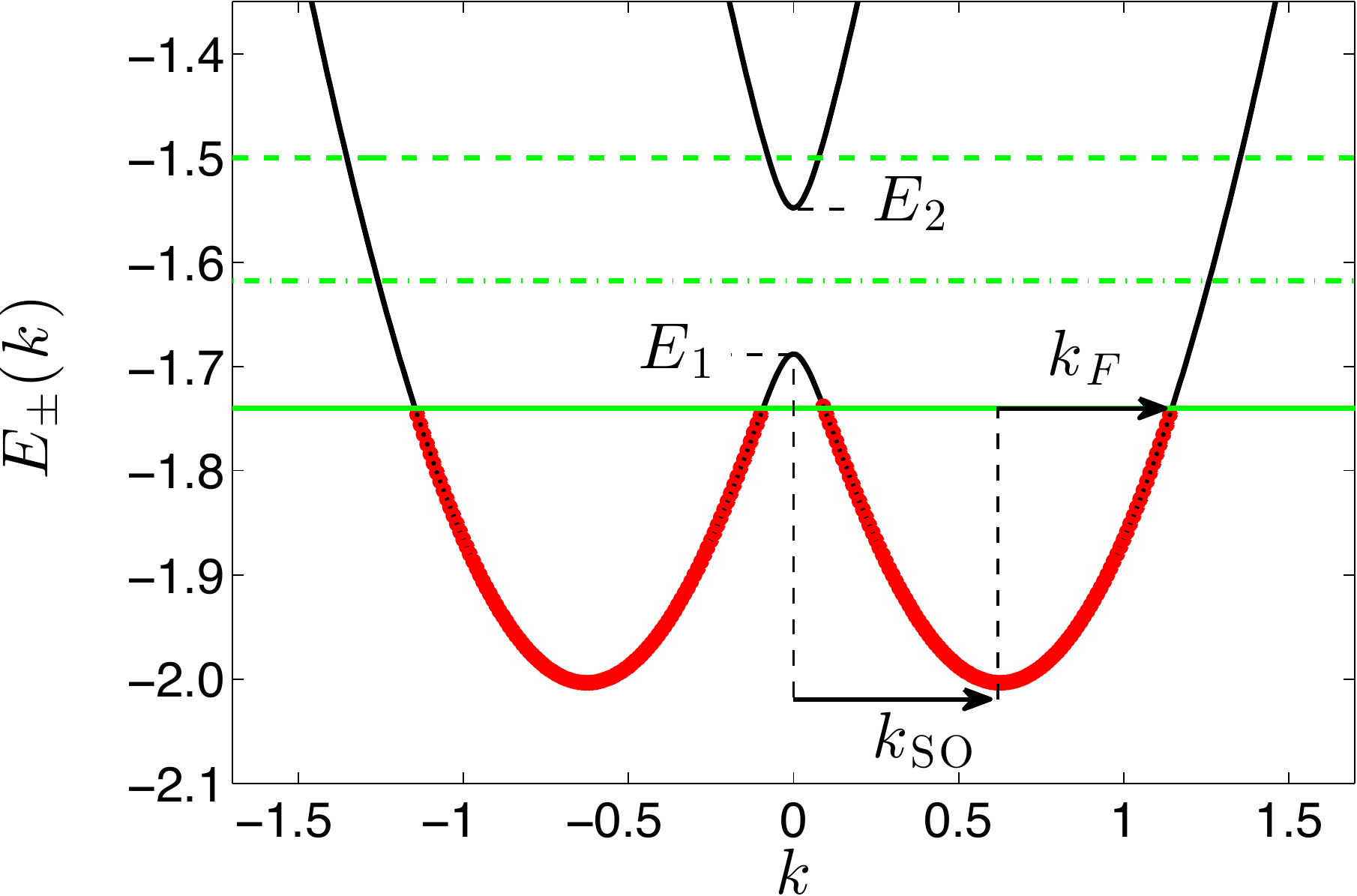}
\caption{(Color online) Band structure $E_\pm(k)$ (solid black lines). The Fermi energy is plotted inside the lower gapless region (green solid line, the red points represent filled states), in the middle of the gap (green dash-dotted line), and inside the upper gapless region (green dashed line). The condition for which $k_F=k_{\rm SO}$ defines the $\nu=1$ phase, when the Fermi energy is at midgap (green dash-dotted horizontal line). The band maximum is denoted by $E_1\equiv E_-(0)$.}
\label{fig:currentbandstructureforintegral}
\end{figure}

\begin{figure*}[t]
\includegraphics[width=5.95cm]{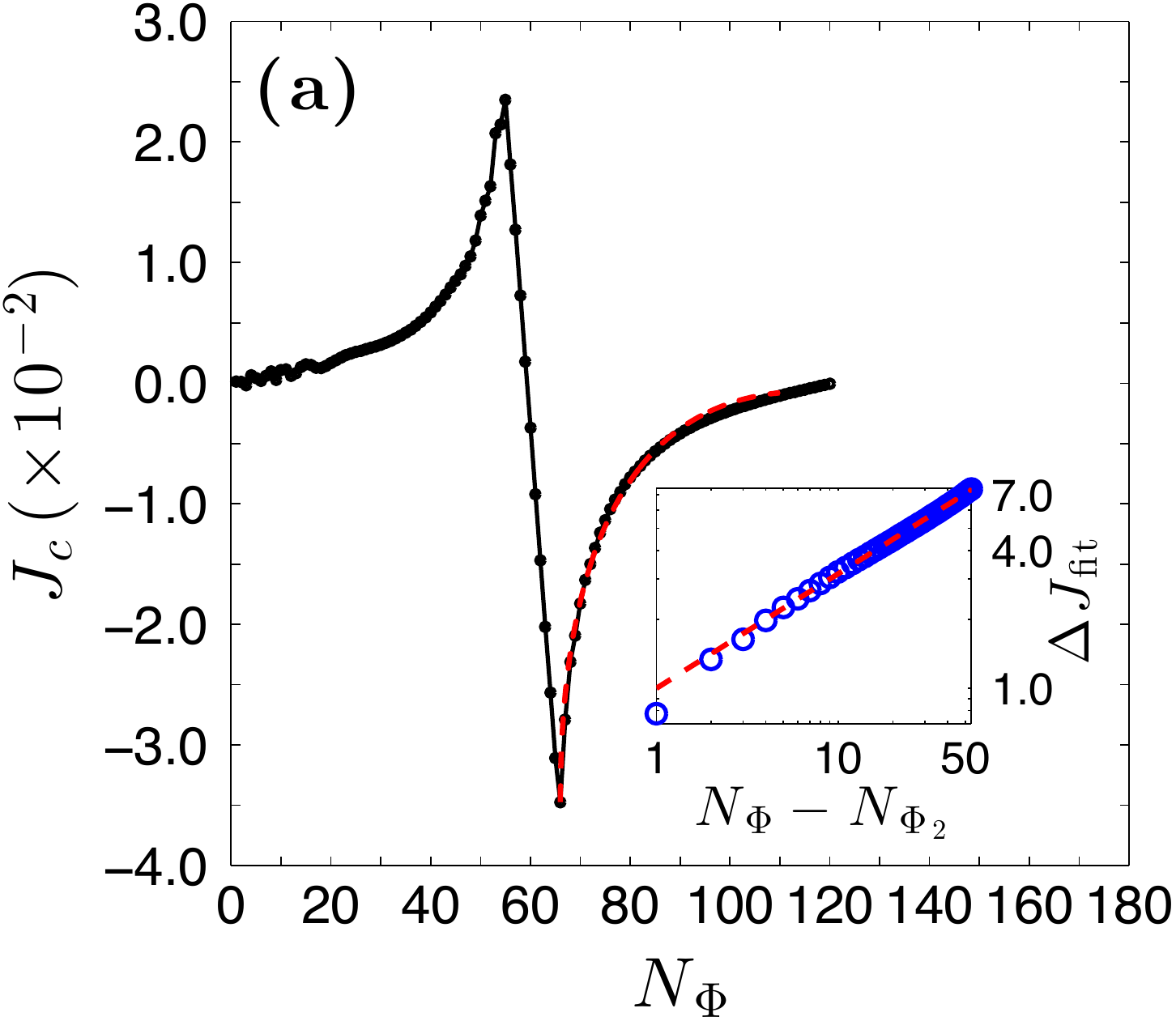}
\includegraphics[width=6.0cm]{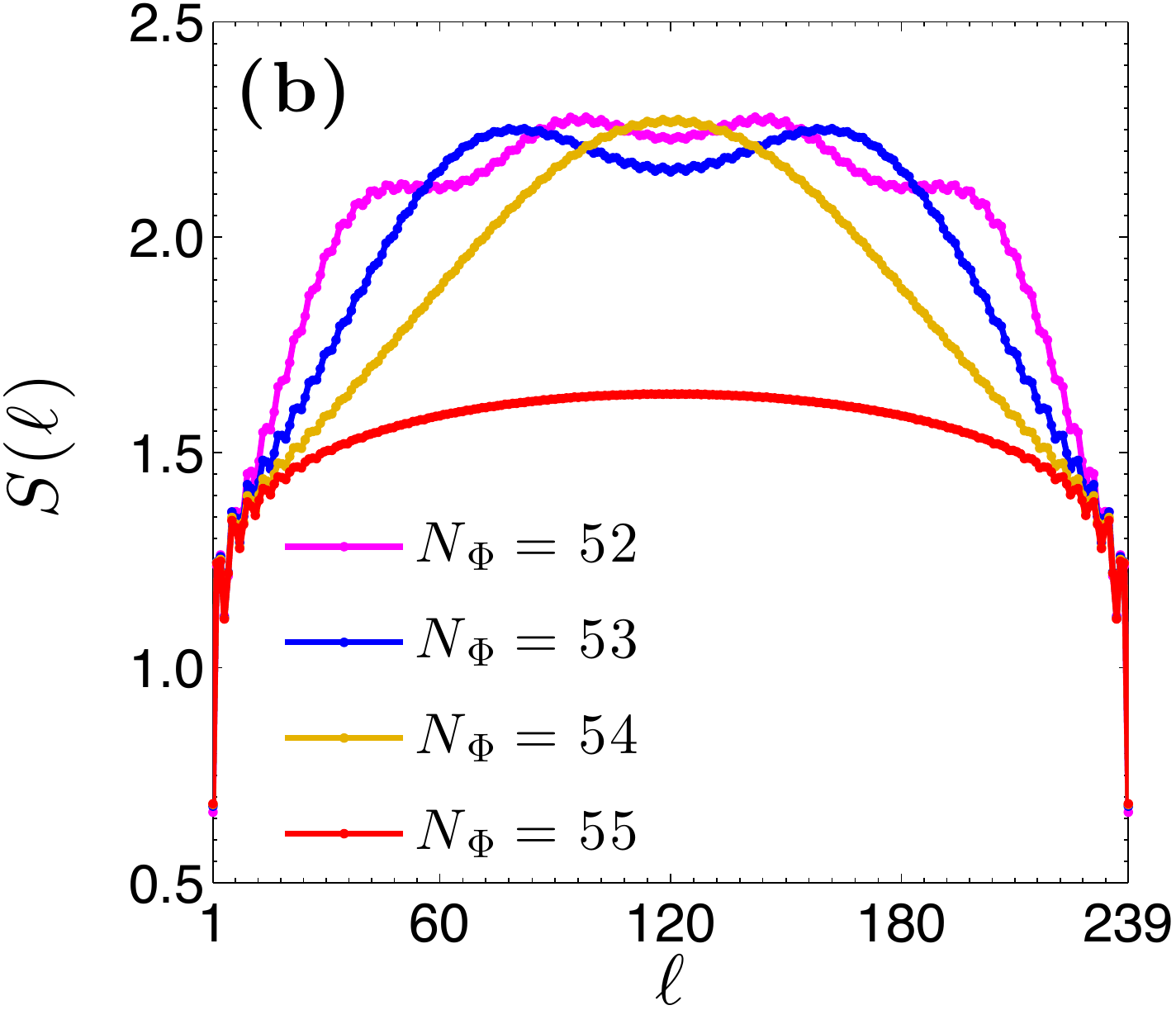}
\includegraphics[width=5.7cm]{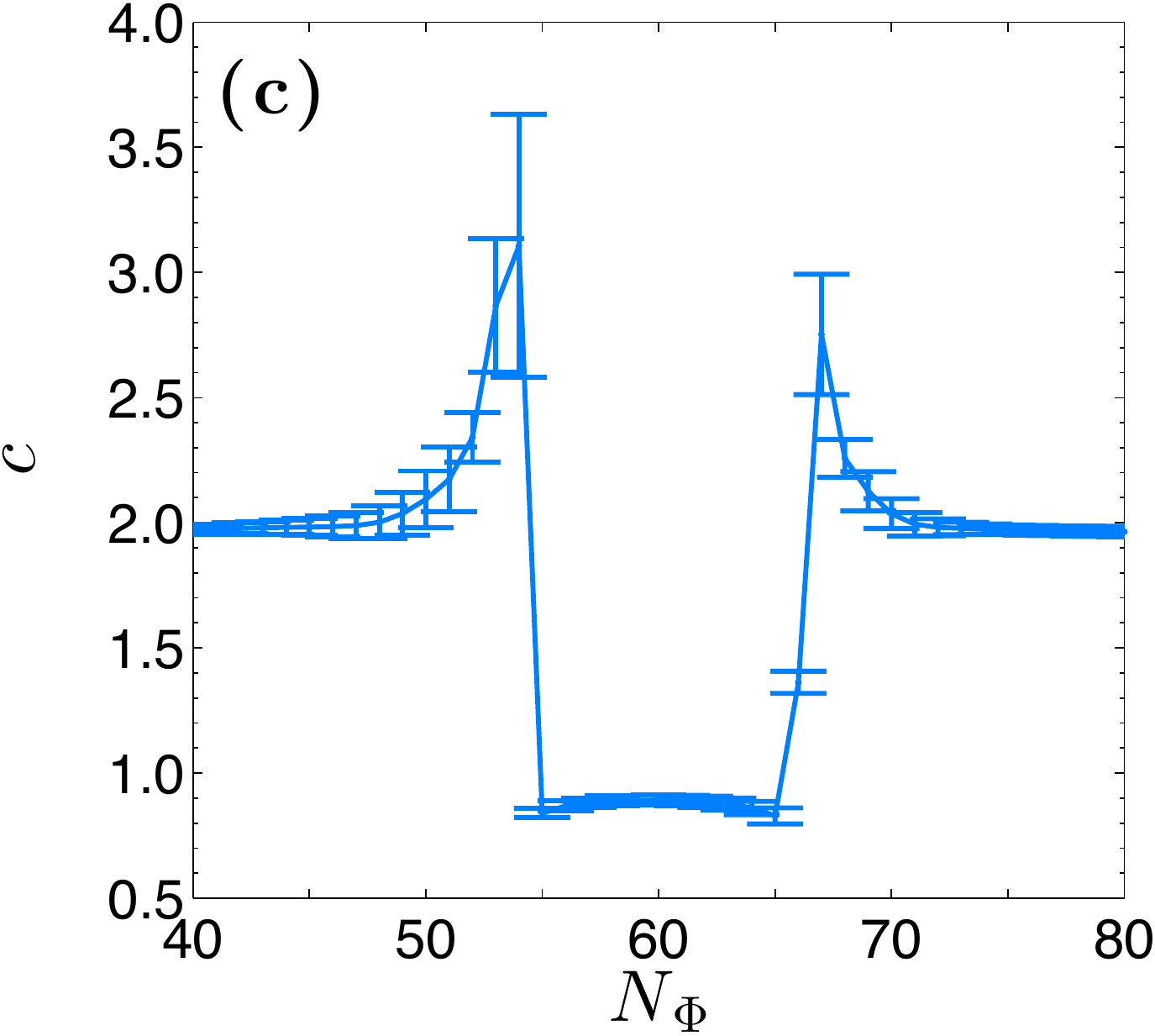}
\caption{(Color online) Numerical results for free fermions ($\nu=1$ state), with $L=240$, $N=120$ and $t_\perp/t=2\times10^{-1}$. We perform the same analysis as in Fig.~\ref{fig:currentcentralchargeandentanglemententropybosons}. The analogy of \textbf{(a)} the chiral current and the fit with Eq.~\eqref{eq:currentsingularbehaviorflux} in the inset, \textbf{(b)} the EE, and \textbf{(c)} the central charge through the C-IC transition with the data on the $\nu=1/2$ Laughlin-like state (Fig.~\ref{fig:currentcentralchargeandentanglemententropybosons}) clearly emerges.}
\label{fig:currentcentralchargeandentanglemententropy}
\end{figure*}

For $t_\perp =0 $, the energy spectrum consists of two shifted cosines, with minima at $k=\pm k_{\rm SO}$, where $\Phi=2k_{\rm SO}$. A nonzero value of $t_\perp$ opens a gap at $k=0,\pm\pi$, where the two bands cross. At zero temperature, the system can thus be in a helical phase, where only two gapless modes with opposite group velocity are found (see Fig.~\ref{fig:currentbandstructureforintegral}). The condition on the fermionic density for this to happen is $k_F=k_{\rm SO}$, where $k_F=\pi n/2$ is the Fermi wave vector. Thus, the filling factor $\nu=2k_F/\Phi=k_F/k_{\rm SO}=1$ identifies this phase~\cite{PhysRevLett.88.036401,PhysRevB.89.085101,PhysRevB.89.115402} (see Fig.~\ref{fig:currentbandstructureforintegral}).

\subsection{Chiral current and integer gap}
\label{sec:chiralcurrentandintegergap}
We now prove that $J_c$, around the transition to the helical phase from a standard non-helical (gapless) phase, exhibits a characteristic cusp singularity, which mirrors the opening of the gap in the energy spectrum at $k=0$. Similar considerations are also presented in Refs.~\cite{PhysRevB.73.195114,PhysRevB.76.195105}.

In order to compute the explicit dependence of $J_c$ on the system parameters and on the Fermi energy, $E_F$, it is convenient to perform the calculation for a fixed spin projection, i.e. $m=+\frac{1}{2}$, the other one being the opposite by symmetry considerations.
In momentum space:
\begin{equation}
\hat J_m=\frac{2t}{L}\sum_k\sin(k-m\Phi)\,\hat a^\dag_{k,m}\hat a_{k,m} \,\, .
\label{eq:currentoperatorcondensedmattergauge}
\end{equation}
Considering only states in the lower band, in the continuum limit, the expectation value of the current operator in Eq.~\eqref{eq:currentoperatorcondensedmattergauge} reads 
$\langle\hat J_m\rangle= \frac{1}{2 \pi}\int dk\,f(k)J_m(k)$, where $f(k)$ is the zero-temperature Fermi-Dirac occupation function, and $J_m(k)=2t\,P^{(k)}_{-,m}\sin(k-m\Phi)$, where $P^{(k)}_{-,m}$ is the probability of finding a fermion at momentum $k$ in the lower band, with spin projection $m$. 

The key result is that the chiral current as a function of the Fermi energy, $E_F$, and for $E_F\sim E_1$, where $E_1\equiv E_-(0)$ (see Fig.~\ref{fig:currentbandstructureforintegral}), is predicted to vary as 
\begin{equation}
J_c\sim\left\{
\begin{array}{ll}
J_{E_1}+A_1\sqrt{E_1-E_F}+A_2(E_1-E_F) &{(E_F<E_1)}\\
\\
J_{E_1}+A_3(E_F-E_1) &{(E_F>E_1)}
\end{array}
\right.
\label{eq:currentsingularbehaviorefermienergy}
\end{equation}
for some coefficients $J_{E_1}$, $A_1$, $A_2$, and $A_3$. We report the details of the derivation of Eq.~\eqref{eq:currentsingularbehaviorefermienergy} in Appendix~\ref{sec:calculationofthecurrentforfreefermions}. 

One can express the current in Eq.~\eqref{eq:currentsingularbehaviorefermienergy} as a function of the particle density $n$ or as a function of the flux $\Phi$. After defining the critical value $n_1$ such that $E_1=E_F(n_1)$, and expanding $E_F(n)-E_1\simeq E'_F(n_1)(n-n_1)$, we obtain:
\begin{equation}
J_c\sim J_{n_1}+B_1\sqrt{n_1-n}+B_2(n_1-n) \quad {(n<n_1)},
\label{eq:currentsingularbehaviordensity}
\end{equation}
for some coefficients $B_1$, $B_2$. A similar argument can be used for the flux dependence: since there exists a critical value $\Phi_1$, at fixed density, such that $E_1=E_F (\Phi_1)$, we similarly obtain
\begin{equation}
J_c\sim J_{\Phi_1}+C_1\sqrt{\Phi_1-\Phi}+C_2(\Phi_1-\Phi) \quad {(\Phi<\Phi_1)} ,
\label{eq:currentsingularbehaviorfluxff}
\end{equation}
for some coefficients $J_{\Phi_1}$, $C_1$ and $C_2$, which is indeed Eq.~\eqref{eq:currentsingularbehaviorflux}. The singular behavior of the current, as in Eqs.~\eqref{eq:currentsingularbehaviordensity} and~\eqref{eq:currentsingularbehaviorfluxff}, thus emerges when the Fermi energy touches the top (or bottom) of a parabolic band, and is related to the so-called van-Hove singularities and signals the occurrence of a Lifshitz transition with dynamical exponent $z=2$.

\subsection{Numerical results}
\label{sec:commensurateincommensuratetransition}

We now present some numerical results for the free-fermion model, to further establish all previous predictions. The goal is to elucidate the analogy between the behavior of the chiral current, central charge and EE we found for the $\nu=1/2$ (see Fig.~\ref{fig:currentcentralchargeandentanglemententropybosons}) and the $\nu=1$ Laughlin-like states.

We simulate the Hamiltonian $\hat H=\hat H_0+\hat H_\perp$ with OBC [see Eq.~\eqref{eq:hamiltonianfermionicsystem}] and measure $J_c$, the central charge and the EE; numerical results are shown in Fig.~\ref{fig:currentcentralchargeandentanglemententropy}. For the parameters we choose, we expect the $\nu=1$ signal at $N_\Phi\simeq N/2=60$. The fit with Eq.~\eqref{eq:currentsingularbehaviorfluxff} confirms the cusp-like behavior after the helical region (Fig.~\ref{fig:currentcentralchargeandentanglemententropy}\textbf{a}).
The values of $c$ that we fit in the helical region (Fig.~\ref{fig:currentcentralchargeandentanglemententropy}\textbf{c}) well agree with $c=1$, whereas outside it we obtain $c=2$, as expected. We ascribe the fact that we do not exactly fit $c=2$ or $c=1$ to the finite value of $L$. The presence of Lifshitz oscillations in the EE is also observed (Fig.~\ref{fig:currentcentralchargeandentanglemententropy}\textbf{b}).

\section{Signatures of Laughlin-like states based on the current}
\label{sec:signaturesbasedonthecurrent}

The results obtained so far pose the following question: is the two-cusp pattern of the current a common feature of Laughlin-like states, related to the occurrence of the Lifshitz transition between a commensurate Laughlin-like phase and an incommensurate standard non-helical phase?

\begin{figure}[t]
\centering
\includegraphics[width=7.7cm]{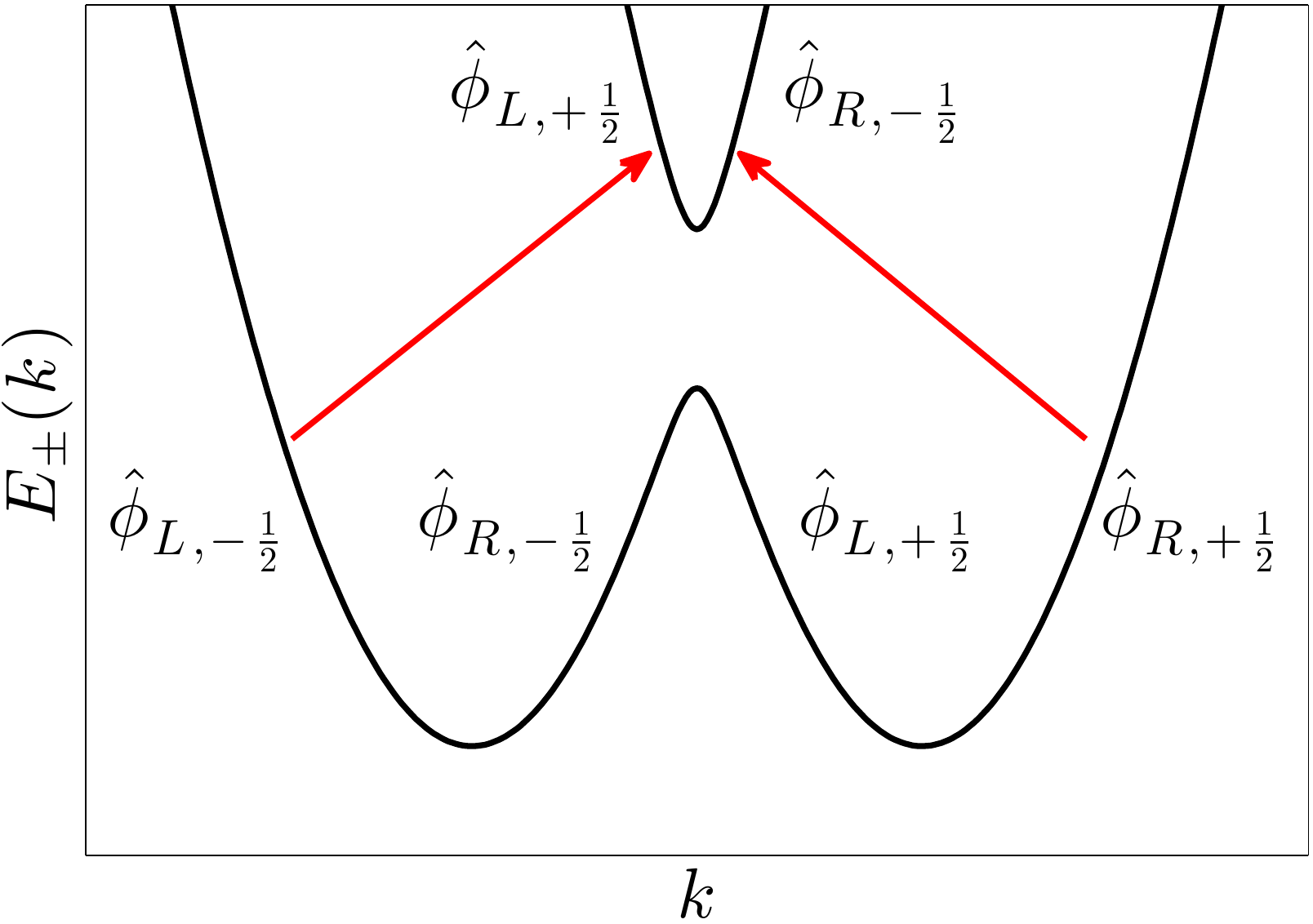}
\caption{(Color online) Band structure $E_{\pm}(k)$ (black lines), and corresponding chiral fields, $\hat \phi_{\mu,m}$ (see Appendix~\ref{sec:bosonizationconventions}). Density-density interaction are $\hat\psi^\dag_{\mu,m}\hat\psi_{\mu,m}\hat\psi^\dag_{\mu',m'}\hat\psi_{\mu',m'}$, where $\mu,\mu'=R,L$ and $m\neq m'$. The pair tunneling process $\hat \psi_{L,-\frac{1}{2}}\hat\psi^\dag_{L,+\frac{1}{2}}\hat\psi^\dag_{R,-\frac{1}{2}}\hat\psi_{R,+\frac{1}{2}}$ is represented by red arrows.}
\label{fig:fermienergybandtunnelingprocess}
\end{figure}

To answer this question, we need to understand if the presence of interactions can affect the dynamical exponent of the C-IC transition. It is known that, when the energy spectrum consists of one parabolic band only, the dynamical exponent, $z$, is unaffected by the presence of interactions~\cite{giamarchi2003quantum}. A different situation arises when the energy spectrum consists of two bands as in our case (see Fig.~\ref{fig:fermienergybandtunnelingprocess}); here, interactions will account for both density-density and pair-tunneling processes between the different branches of the dispersion. As discussed in Refs.~\cite{PhysRevLett.98.126404, PhysRevLett.102.176404}, pair-tunneling processes can change the universality class of the Lifshitz transition. Following Refs.~\cite{PhysRevLett.98.126404, PhysRevLett.102.176404}, it is possible to understand whether pair tunneling affects the Lifshitz transition by combining bosonization and renormalization group (RG) techniques.

\subsection{Square-root scaling via RG approach}
Here, we argue that the square-root singularity in the double-cusp structure is preserved also when interactions are present, and hence serves as a signature for Laughlin-like states.
In bosonization language (see Appendix~\ref{sec:bosonizationconventions}), the Hamiltonian in Eqs.~\eqref{eq:hamiltonianhardcorebosons} and~\eqref{eq:hamiltonianfermionicsystem} reads:
\begin{eqnarray}
\hat H&=&\sum_{\lambda=c,s}\frac{u_\lambda}{2}\int dx \left[\frac{1}{K_\lambda}{\left(\partial_x\hat \varphi_\lambda\right)}^2+K_\lambda{\left(\partial_x\hat \theta_\lambda\right)}^2\right]\nonumber\\
\nonumber\\
&&+A_{\rm pair} g_{\rm pair}\int dx \cos\left(2\sqrt{2\pi}\,\hat \varphi_s\right)\nonumber\\
\nonumber\\
&&+A_{\rm FQH}\, g_{\rm FQH}\int dx \cos\left[\sqrt{2\pi}\left(\hat\theta_s+p\,\hat \varphi_c\right)+\delta x\right] \,\, , \nonumber\\
\label{eq:RG1}
\end{eqnarray}
where $A_{\rm pair}$ and $A_{\rm FQH}$ are cutoff-dependent constants, and $p$ is an odd (even) integer for fermions (bosons). The parameter $\delta=\Phi-p\,\pi n$ represents the deviation from commensurability in the $\nu=1/p$ FQHE. Density-density interactions are encoded in the Luttinger parameters, $K_c$ and $K_s$, in Eq.~\eqref{eq:RG1}. As shown in Refs.~\cite{PhysRevLett.98.126404,PhysRevLett.102.176404}, these density-density interactions do not change the nature of the transition. The pair-tunneling process is identified by the operator $\hat \psi_{L,-\frac{1}{2}}\hat\psi^\dag_{L,+\frac{1}{2}}\hat\psi^\dag_{R,-\frac{1}{2}}\hat\psi_{R,+\frac{1}{2}}$. In bosonization language, this operator corresponds to the term $\cos(2 \sqrt{2 \pi}\,\hat\varphi_s)$ in Eq.~\eqref{eq:RG1}. The presence of such a term, depending on its relevancy under RG, may change the universality class of the Lifshitz transition.

\begin{figure*}[t]
\centering
\includegraphics[width=5.9cm]{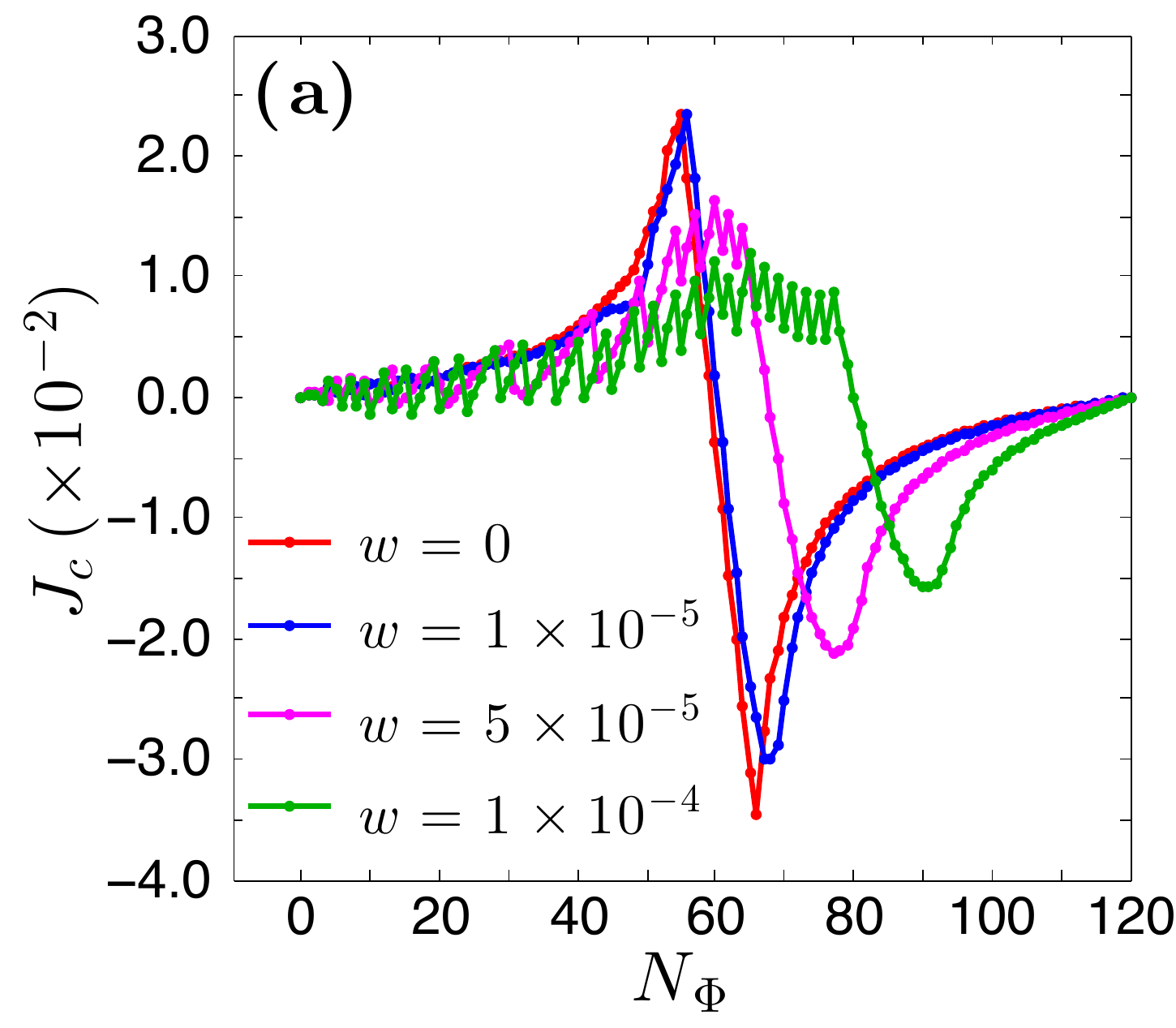}
\includegraphics[width=5.9cm]{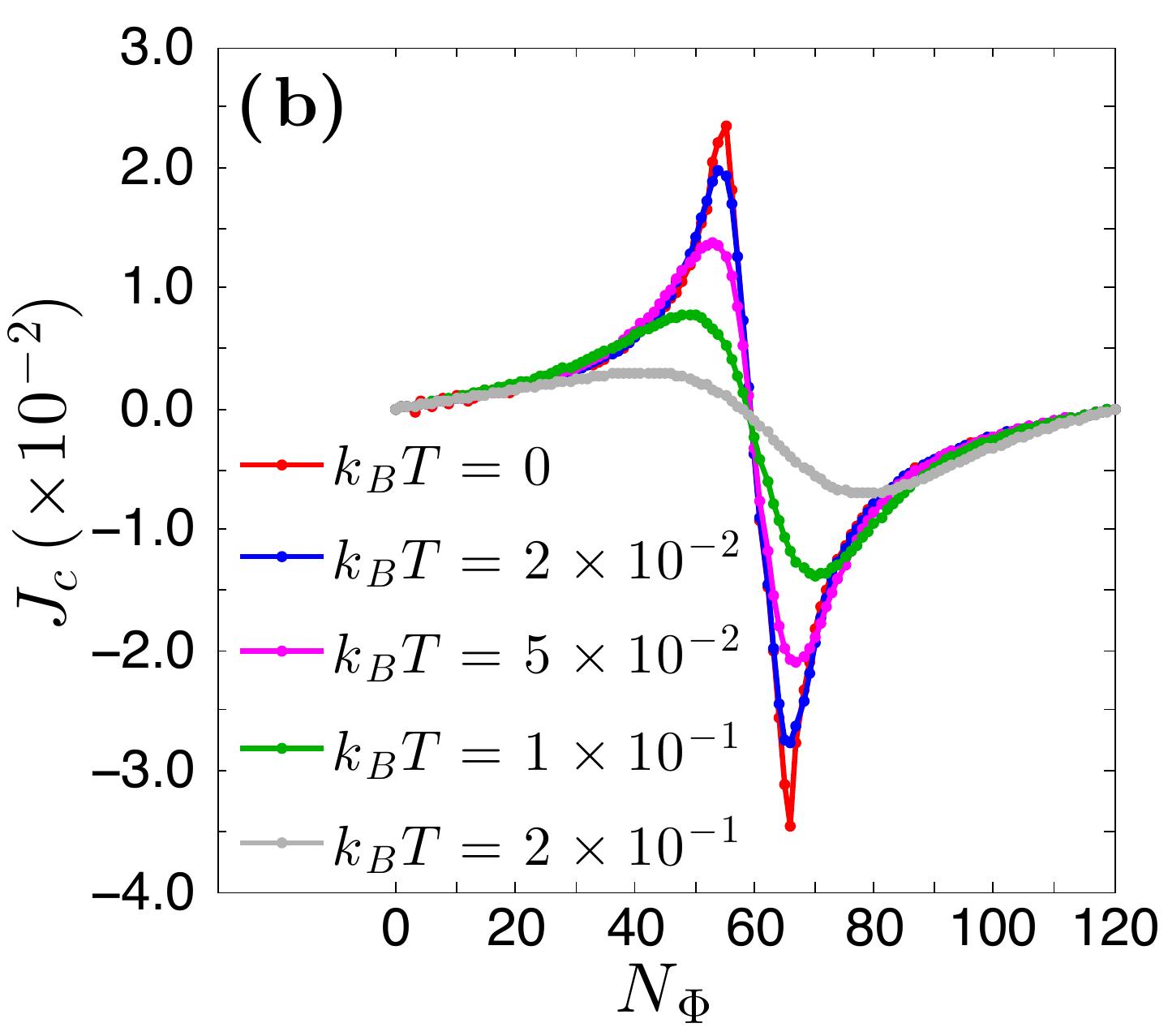}
\includegraphics[width=5.93cm]{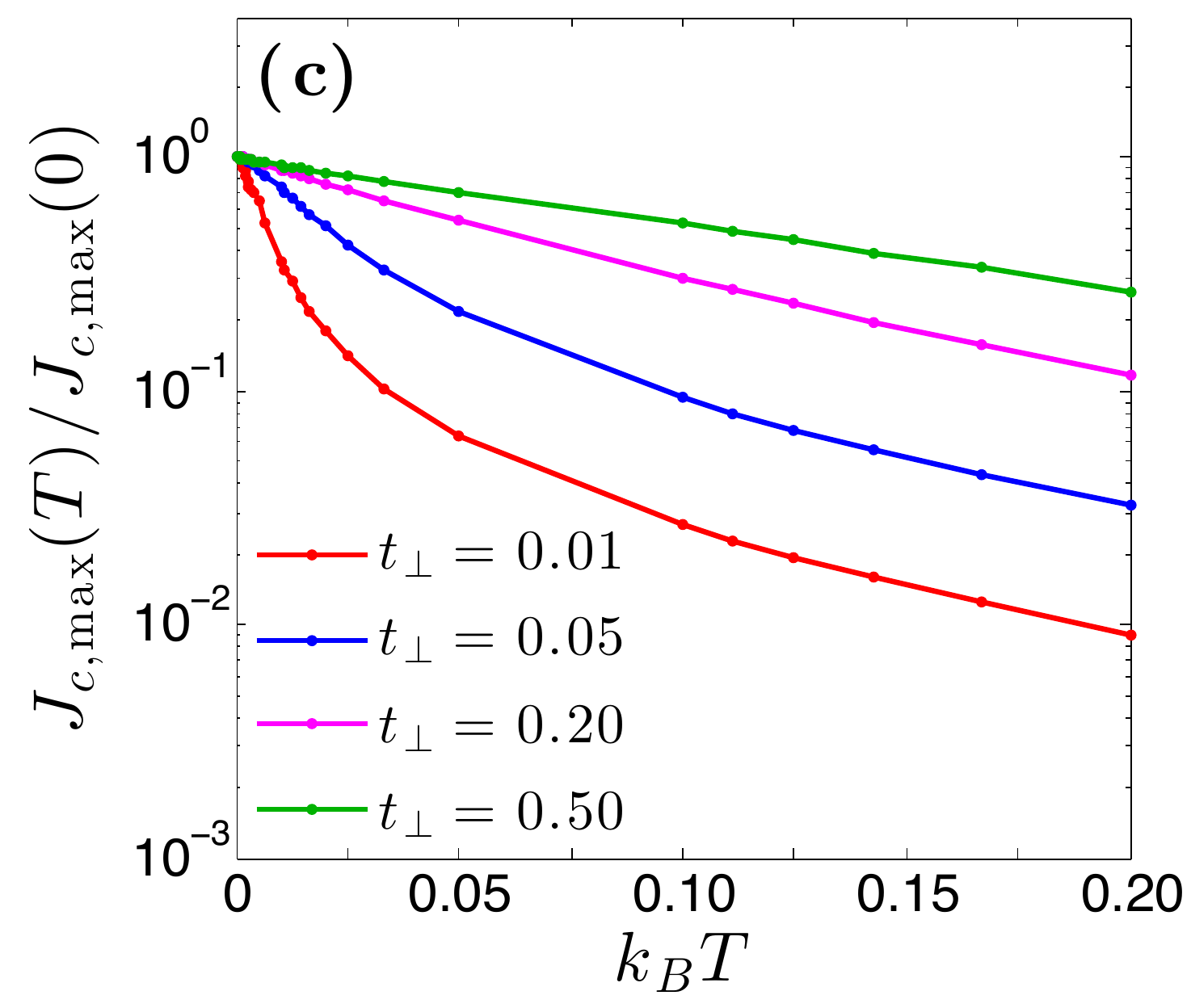}
\caption{{(Color online) Numerical results for free fermions ($\nu=1$ Laughlin-like state). \textbf{(a)} Simulation with $L=240$, $N=120$ and $t_\perp/t=2\times10^{-1}$, as in Fig.~\ref{fig:currentcentralchargeandentanglemententropy}, with the inclusion of a finite harmonic trap, for different values of $w$. The harmonic trap is implemented as explained in the text. \textbf{(b)} Simulations at finite temperature. We simulate free fermions with $L=240$, $N=120$ and $t_\perp/t=2\times10^{-1}$ for different values of $k_BT$. \textbf{(c)} We show the ratio between the peak of $J_c$ at temperature $T$ and the peak at zero temperature. To increase numerical accuracy, we simulate free fermions with $L=480$, $N=240$, and with different values of $t_\perp/t$.}}
\label{fig:figureatfiniteomegaandtemperature}
\end{figure*}

We now perform a RG analysis for the Hamiltonian Eq.~\eqref{eq:RG1}, away from the commensurability condition i.e. $\delta\not\approx0$, and starting at small coupling $g_{\rm FQH}\ll~1$, $K_s\approx~1$. Following Ref.~\cite{giamarchi2003quantum}, we write the following RG equations:
\begin{subequations}
\begin{align}
\frac{dg_{\rm pair}}{dl} &= (2-\Delta_{\rm pair})g_{\rm pair} \,\,,\,\, \frac{dg_{\rm FQH}}{dl} = (2-\Delta_{\rm FQH})g_{\rm FQH}\,\, ,\\
\frac{dK_s}{dl} &=\frac{1}{8}g_{\rm FQH}^2\,J_0(e^l\delta\Lambda)-\frac{1}{2}g_{\rm pair}^2K_s^2\,\, ,
\end{align}
\label{eq:RG2}
\end{subequations}
where $l$ is the running parameter of the flow, $\Lambda$ is the ultraviolet cutoff of the Luttinger theory (see Appendix~\ref{sec:bosonizationconventions}), $\Delta_{\rm pair}=2K_s$, $\Delta_{\rm FQH}=\frac{1}{2}\left(K_s^{-1}+p^2K_c\right)$, and $J_0(\cdot)$ is the Bessel function of the first kind.

We use the following simple argument, describing the competition between the flow towards strong coupling of $g_{\rm pair}$ versus $g_{\rm FQH}$. For sufficiently small $K_c$, which we assume for the realization of FQH-like states, we see from Eqs.~\eqref{eq:RG2} that the $g_{\rm FQH}$ cosine is relevant, and it increases $K_s$ and $\Delta_{\rm pair}$, hence making the $g_{\rm pair}$ cosine irrelevant. Thus, the $g_{\rm pair}$ cosine will be irrelevant, and the cusp behavior near the phase transition will thus signal the Lifshitz C-IC transition, with $z=2$. Specifically, this perturbative analysis is better controlled if one adds weak interactions on top of an integer $\nu=1$ state, in which case $K_s, K_c \to 1$, and $g_{\rm pair}$ is small.

\subsection{Square-root scaling via refermionization}
The previous RG analysis allows us to write the Hamiltonian in Eq.~\eqref{eq:RG1} without the $g_{\rm pair}$ term. One can turn to the rescaled fields by means of the canonical transformation
$\tilde{\varphi}_c=p\,\hat \varphi_c$, $\tilde{\theta}_c=\hat \theta_c/p$, $\tilde{\varphi}_s=\hat \varphi_s$ and $\tilde{\theta}_s=\hat \theta_s$, and obtain the Hamiltonian
\begin{eqnarray}
\hat H&=&\frac{u_s}{2}\int dx\,\left[K_s{\left(\partial_x\tilde{\theta}_s\right)}^2+\frac{1}{K_s}{\left(\partial_x\tilde{\varphi}_s\right)}^2\right]\nonumber\\
\nonumber\\
&&+\frac{u_c}{2}\int dx\,\left[p^2K_c{\left(\partial_x\tilde{\theta}_c\right)}^2+\frac{1}{p^2K_c}{\left(\partial_x\tilde{\varphi}_c\right)}^2\right]\nonumber\\
\nonumber\\
&&+A_{\rm FQH}\,g_{\rm FQH}\int dx\,\cos\left[\sqrt{2\pi}\left(\tilde{\theta}_s+\tilde{\varphi}_c\right)\right] \,\, .
\label{eq:hamiltonianlaughlinrescaled}
\end{eqnarray}
If the interaction parameters of the original microscopic model are such that the initial value of the Luttinger parameter is $K_c=1/p^2$ and $u_c=u_s$ (what we refer to as the refermionization point) one recovers, for an SU(2) invariant system, an effective Hamiltonian with Luttinger parameters equal to one, i.e. a free theory in terms of the fields $\tilde\varphi_c$, $\tilde\theta_c$, $\tilde\varphi_s$, and $\tilde\theta_s$. The refermionization procedure~\cite{PhysRevLett.33.589,PhysRevB.89.085101} thus allows one to define new fermionic operators $\tilde \psi_m$, for $m=\pm1/2$, and their corresponding discretized version on a lattice, $\tilde a_{j,m}$ ($\tilde a^\dag_{j,m}$). For the refermionized system, the picture seen in Sec.~\ref{sec:IQHE} holds. Because of the $\hat\varphi_c$ rescaling, the new particle density is $\tilde n=p\,\hat n$.

We see that, at the refermionization point, $\Delta_{\rm FQH}=1$, and the gap scales as $\Delta E_{\rm FQH}\sim(A_{\rm FQH}\,g_{\rm FQH}/t)$, as expected from free fermions. The refermionized Hamiltonian is that of Eq.~\eqref{eq:hamiltonianfermionicsystem}, with $U=V=0$, in terms of the $\tilde a_{j,m}$ and $\tilde a^\dag_{j,m}$ operators, and with transverse hopping $A_{\rm FQH}\,g_{\rm FQH}$. Then, using the results from Sec.~\ref{sec:IQHE}, the chiral current exhibits the singular behavior in Eq.~\eqref{eq:currentsingularbehaviordensity} and Eq.~\eqref{eq:currentsingularbehaviorfluxff}, but with critical values, $n_{1,2}$ and $\Phi_{1,2}$, which will in general depend on $p$.

We stress that the reaching of the refermionization point requires an extreme fine tuning of the Luttinger parameters and sound velocities which, in general, can not be varied independently. However, the analysis of the previous section can be applied to study small deviations from the refermionized points, showing that the nature of the Lifshitz transition remains intact.

{\section{Experimental realizability}}
\label{sec:experimentalfeasibility}
{In this section, we discuss the experimental relevance of our systems. The realization of two-leg ladders and the observation of chiral edge currents have been recently reported using alkaline and alkaline-earth-like atoms both in the fermionic and in the bosonic case~\cite{science1510,science1514}. Let us also mention the proposal in Ref.~\cite{njp.12.033007}, where the two-leg ladder is realized by using the proper bosonic or fermionic isotope of Yb, and by representing the two legs of the ladder with the GS ($1S^0$) and the long-lived metastable excited state ($3P^0$), as recently reported in Refs.~\cite{PhysRevLett.117.220401,10.1038.nature20811}.}

{Precise measurements of the particle density at the level of the single site, which are needed to ensure the commensurability condition between $n$ and $\Phi$ required by the Laughlin-like states, are nowadays experimentally accessible by using single-site-resolved imaging, both for fermionic~\cite{natphys11.738,PhysRevLett.115.263001,PhysRevA.92.063406,PhysRevLett.114.193001,PhysRevLett.114.213002} and bosonic gases~\cite{nature462.7477,nature467.68}.}

{In our numerical simulations, in the situation in which the double-cusp pattern of $J_c$ is preserved, the chiral current takes values which are of the order of $10^{-3}\,t$ or $10^{-2}\,t$ (see Fig.~\ref{fig:currenttperpandvperpscan}). Such values lay at the edge of present experimental sensitivity. Indeed, in Ref.~\cite{NatPhys10.588.14} measurements of $J_c$ in the bosonic ladder of the order of $10^{-2}\,t$ are reported.}

{Because of the smallness of the signal, we need to discuss the stability of such a small signature in the presence of a trapping potential and finite temperature $T$. To study the effects of a trap and of finite temperature, we focus on the fermionic $\nu=1$ Laughlin-like state, which is the easiest case we can numerically study, and it is expected to provide the correct qualitative behavior. The harmonic trap is described by the Hamiltonian $\hat H_{\rm trap}=\sum_{m=\pm1/2}\sum_jw{(j-L/2-1/2)}^2\,\hat a^\dag_{j,m}\hat a_{j,m}$. We compute $J_c$ for different $w$ at $T=0$ (Fig.~\ref{fig:figureatfiniteomegaandtemperature}\textbf{a}), using $L=240$ and $N=120$ as in Fig.~\ref{fig:currentcentralchargeandentanglemententropy}. From our simulations, we see that the presence of an harmonic trapping potential hinders the possibility of detecting the Laughlin-like state because the double-cusp pattern of $J_c$ is lost even for small $w$. This makes the use of a hard-wall box potential necessary, whose experimental realization has been reported in Refs.~\cite{PhysRevLett.110.200406,PhysRevLett.113.135302,ncomms.6.6172}, and mimics the open boundary conditions we consider in our simulations.}

{We then study the effect of finite temperature $T$ simulating free fermions at $w=0$. We show in Fig.~\ref{fig:figureatfiniteomegaandtemperature}\textbf{b} the chiral current using the same parameters in Fig.~\ref{fig:currentcentralchargeandentanglemententropy}, for different values of $T$. In Fig.~\ref{fig:figureatfiniteomegaandtemperature}\textbf{c}, we show the ratio $J_{c,\rm max}(T)/J_{c,\rm max}(0)$, where $J_{c,\max}(T)$ is the value of the peak of $J_c$ at the boundary between the non-helical and the helical region, for different values of $t_\perp/t$. We see that, for the values of $t_\perp/t$ used in Fig.~\ref{fig:figureatfiniteomegaandtemperature}\textbf{b} and Fig.~\ref{fig:figureatfiniteomegaandtemperature}\textbf{c}, the double cusp of $J_c$ is still visible (e.g., reduced by less than one order of magnitude) for $k_BT\sim 0.1\,t$, where $k_B$ is the Boltzmann constant, and thus still experimentally measurable. For larger temperature, the two-cusp pattern of the chiral current is suppressed.}

%%%

\vspace{0.8cm}
\section{Conclusions and perspectives}
\label{sec:conclusions}

In this article we have presented extensive numerical evidence of Laughlin-like states in experimentally-relevant models of bosonic and fermionic gases with a synthetic dimension.
Our characterization rests on several signatures based on the chiral current and on the entanglement entropy of the ground state.
By providing a first numerical viewpoint on the study of the quantum Hall effect in coupled arrays of one-dimensional systems, our study opens a number of interesting perspectives.

From a theoretical point of view, it is particularly important to move away from the weak-coupling limit $t_\perp/t \ll 1$ where bosonization is usually performed, and to understand the nature of Laughlin-like states when $t_\perp/t \sim 1$. In this respect, a characterization of the fractional conductance which unambiguously defines the Laughlin-like state in an equilibrium setting might be crucial.
On the other hand, the one-dimensional nature of these setups prevents us from speaking of a truly topologically-protected phase, but it is still to be investigated which properties are inherited from the two-dimensional case, focusing in particular on the entanglement spectrum and on the nature of the excitations.

The difficulty of the numerical problem, reflected by an extremely low convergence rate of the variational procedure of our MPS algorithms, has prevented us from addressing in a more systematic way the whole series of Laughlin-like states. 
The elaboration of ad-hoc variational methods and the formulation of models with finite-range interactions where Laughlin-like states could be more robust is an important perspective. The extension of this study to gases with higher spin, and thus longer synthetic dimension, is also desirable.

From an experimental point of view, our study clearly points toward the possibility of using synthetic dimension approaches to investigate quantum Hall physics in ultra-cold gases. Our {results can be} immediately relevant to experiments using Yb atoms in optical lattices, where {the role of the two legs of the ladder is played by internal electronic states (the case of GS $1S^0$ and metastable excited state $3P^0$ has been recently reported in Refs.~\cite{PhysRevLett.117.220401,10.1038.nature20811})}. {The identification of Laughlin-like states should be performed through time-of-flight measurements of the current flowing into the system~\cite{science1510}; the small expected signal poses some constraints on the experimental sensitivity, since values smaller than $\sim10^{-2}\,t$ have to be resolved. Furthermore, since the two-cusp pattern which identifies such phase disappears even in the presence of a tiny harmonic confinement, the implementation of a box potential is necessary. Notwithstanding these significant challenges,} our findings {show a viable} route to the demonstration of Laughlin-like states - a fundamental step along the way of understanding the combined effect of interaction and static gauge potentials in atomic physics experiments.

\textit{Note added}: Recently, we became aware of a related study performed by Petrescu \textit{et al.}~\cite{arXiv:1612.05134}.

\acknowledgments

We acknowledge fruitful discussions with P.~Azaria, M.~Burrello, M.~Garst, T.~Giamarchi, A.~Haller, F.~Heidrich-Meisner, M.~Piraud, M.~Rizzi, G.~Roux and L.~Taddia. {We are grateful to F.~Stra for technical support.} R.~F.~was supported through EU project QUIC. L.~M.~was supported by LabEX ENS-ICFP: ANR-10-LABX-0010/ANR-10-IDEX-0001-02 PSL*. E.~S. was supported by ISF Grant No. 1243/13, and the Marie Curie CIG Grant No. 618188 (ES). We acknowledge the CINECA award under the ISCRA initiative, for the availability of high performance computing resources and support.

\appendix

\section{Bosonization analysis}
\label{sec:bosonizationconventions}
In this appendix, we recall the bosonization analysis for Laughlin-like states, following Refs.~\cite{PhysRevLett.88.036401,PhysRevB.89.085101,PhysRevB.91.054520,PhysRevB.92.115446}. In the continuum limit, one can express the fermionic operator $\hat a_{j,m}$ as a field operator $\hat\psi_m(x)$, and describe the low-energy physics in terms of the four chiral fields
\begin{equation}
\hat\psi_m(x)\simeq\hat\psi_{R,m}(x)+\hat\psi_{L,m}(x) \,\, ,
\label{eq:fermionicfieldleftandrightdecomposition}
\end{equation}
where $\{\hat\psi_{\mu,m}\}_{\mu=R,L}$ represent the \emph{right} and \emph{left moving} fields, for $m=\pm1/2$. Fermionic fields are written in the bosonic representation
\begin{eqnarray}
\hat\psi_{R,L;m}(x)&=&\frac{e^{\mp i\pi\frac{n}{2}x}}{\sqrt{2\pi\Lambda}}\,e^{-i\sqrt{2\pi}\,\hat\phi_{R,L;m}(x)} \nonumber\\
\nonumber\\
&=&\frac{e^{\mp i\pi\frac{n}{2}x}}{\sqrt{2\pi\Lambda}}\,e^{-i\sqrt{\pi}\,\left[\hat\theta_m(x)\mp\hat\varphi_m(x)\right]} \,\, ,
\label{eq:fermionicfieldbosonicrepresentation}
\end{eqnarray}
where $n=\frac{1}{L}\sum_jn_j=\frac{1}{L}\sum_j\sum_m\langle\hat a^\dag_{j,m}\hat a_{j,m}\rangle$ is the total fermionic density, $\Lambda$ is the ultraviolet momentum cutoff of the theory, and $\{\hat\phi_{\mu,m}\}$ are the four \emph{chiral fields}, describing the low-energy bosonic excitations about the Fermi surface. The $\hat\theta_{\pm1/2}$ and $\hat\varphi_{\pm1/2}$ are the \emph{phase} and \emph{density} dual fields respectively:
\begin{equation}
\hat\phi_{\mu,m}=\frac{1}{\sqrt{2}}\left(\hat\theta_m-\mu\,\hat\varphi_m\right) \,\, .
\end{equation}
where $\mu=\pm$ for $R,L$. In the thermodynamic limit ($L\rightarrow\infty$), dual fields obey the canonical commutation relations (CCR) $[\hat\varphi_{m}(x),\partial_y\hat\theta_{m'}(y)]=i\,\delta_{m,m'}\,\delta(x-y)$.

It is customary to introduce the dual \emph{charge} and \emph{spin} fields $\hat\varphi_{c,s}$ and $\hat\theta_{c,s}$, such that
\begin{equation}
\hat\phi_{\mu,m}=\frac{1}{2}\left(\hat\theta_c-\mu\,\hat\varphi_c+ {\rm sgn}(m)\,\hat\theta_s-{\rm sgn}(m)\mu\,\hat\varphi_s\right) \,\, ,
\label{eq:chiralfieldsintermsofchargeandspinfields}
\end{equation}
where $\mu=\pm$ for $R,L$ respectively. Also the $\hat\varphi_{c,s}$ and $\hat\theta_{c,s}$, fields obey the CCR, which read as $[\hat\varphi_{\lambda}(x),\partial_y\hat\theta_{\lambda'}(y)]=i\,\delta_{\lambda,\lambda'}\,\delta(x-y)$. In the bosonic representation, Hamiltonian in Eq.~\eqref{eq:hamiltonianfermionicsystem}, with $U=V=0$, is the standard Luttinger liquid Hamiltonian, plus the sine-Gordon Hamiltonian describing transverse hopping, and reads:
\begin{eqnarray}
\hat H&=&\frac{v_F}{2}\sum_{\lambda=c,s}\int dx\,\left[{\left(\partial_x\hat\varphi_\lambda\right)}^2+{\left(\partial_x\hat\theta_\lambda\right)}^2\right]\nonumber\\
\nonumber\\
&&+\frac{t_\perp}{\pi\Lambda}\int dx\,\cos\left[\sqrt{2\pi}\left(\hat\varphi_c+\hat\theta_s\right)+(\Phi-\pi n)x\right] \,\, , \nonumber\\
\label{eq:bosonizationhamiltonianfreefermionschargeandspin}
\end{eqnarray}
where $v_F$ is the Fermi velocity. In deriving Hamiltonian~\eqref{eq:bosonizationhamiltonianfreefermionschargeandspin}, only potentially non-oscillating terms have been retained. The cosine term in Eq.~\eqref{eq:bosonizationhamiltonianfreefermionschargeandspin} is oscillating unless $n=\Phi/\pi$, which is the resonance condition for the $\nu=\frac{\pi n}{\Phi}=1$ phase to occur.

In the charge and spin basis, interactions are taken into account in the Luttinger liquid Hamiltonian by introducing the Luttinger parameters, $K_c$ and $K_s$, and the charge and spin sound velocities, $u_c$ and $u_s$, for both spin and charge sectors. Thus, when interactions are considered, the Luttinger liquid part in Eq.~\eqref{eq:bosonizationhamiltonianfreefermionschargeandspin} becomes
\begin{equation}
\hat H^{(LL)}=\sum_{\lambda=c,s}\frac{u_\lambda}{2}\int dx\,\left[\frac{1}{K_\lambda}{\left(\partial_x\hat\varphi_\lambda\right)}^2+K_\lambda{\left(\partial_x\hat\theta_\lambda\right)}^2\right] \,\, .
\label{eq:hamiltonianinteractionluttingerliquid}
\end{equation}
For free fermions and HCBs, $K_c=K_s=1$. If interactions do not break SU(2) symmetry, $K_s=1$ also in presence of interactions, and the effect of interactions is written in $K_c$. For repulsive interaction, $K_c<1$. In the limit of infinite on-site interaction, one has $K_c=1/2$. With long-range interaction, one can achieve $K_c<1/2$.

Furthermore, because of band curvature effects, the decomposition of the fermionic field, given by Eq.~\eqref{eq:fermionicfieldbosonicrepresentation}, has to be modified by including all harmonics~\cite{PhysRevLett.47.1840}
\begin{equation}
\hat\psi_m(x)=\sum_{p}\alpha_p\,e^{-ip\pi\frac{n}{2}x}\,e^{-i\sqrt{\pi}\left[\hat\theta_m(x)-p\,\hat\varphi_m(x)\right]} \,\, ,
\label{eq:fieldallharmonic}
\end{equation}
where, in Eq.~\eqref{eq:fieldallharmonic}, $p$ is an odd (even) number for fermionic (bosonic) fields, and $\{\alpha_p\}$ are some non-universal expansion coefficients. Hamiltonian in Eq.~\eqref{eq:hamiltonianlatticenogaugeshift} becomes
\begin{equation}
\hat H_\perp=\sum_pt_{\perp,p}\int dx\,\cos\left[\sqrt{2\pi}\left(\hat\theta_s+p\,\hat\varphi_c\right)+(\Phi-p\pi n)x\right]
\label{eq:hamiltonianmagneticinteractionlaughlin}
\end{equation}
for some non-universal coefficients $\{t_{\perp,p}\}$. The cosine operator in Eq.~\eqref{eq:hamiltonianmagneticinteractionlaughlin}, which we refer to as the Laughlin operator~\cite{PhysRevLett.88.036401}, for a given $p$, has scaling dimension
\begin{equation}
\Delta_{\rm FQH}=\frac{1}{2}\left(p^2K_c+K^{-1}_s\right) \,\, ,
\label{eq:scalingdimensionlaughlinprocess}
\end{equation}
and is non oscillating if $\Phi=p\pi n$, which yields the Laughlin sequence $\nu=\frac{\pi n}{\Phi}=\frac{1}{p}$. When the fractional gap opens, its magnitude is predicted to scale as
\begin{equation}
\Delta E_{\rm FQH}\sim{\left(\frac{t_\perp}{t}\right)}^{1/(2-\Delta_{\rm FQH})} \,\, ,
\label{eq:gapfromscalingdimension}
\end{equation}
in the deep massive phase and for sufficiently small values of $t_\perp/t$~\cite{giamarchi2003quantum}.

{For the $\nu=1/2$ Laughlin-like state, by using the expansion in Eq.~\eqref{eq:fieldallharmonic}, the transverse current operator in bosonization language, around $2\pi n=\Phi$, reads
\begin{equation}
\hat J_{\perp}(x)\sim\sin\left[\sqrt{2\pi}\left(\hat\theta_s+2\,\hat\varphi_c\right)\right] \,\, .
\label{eq:transversecurrentoperatorbosonization}
\end{equation}
The pinning of the combination of fields $\hat\theta_s+2\,\hat\varphi_c$, when the Laughlin-like state is encountered, implies that $\langle\hat J_\perp(x)\rangle=0$, and that transverse current fluctuations are suppressed in the Laughlin-like region.}

\begin{figure*}[t]
\centering
\includegraphics[width=8cm]{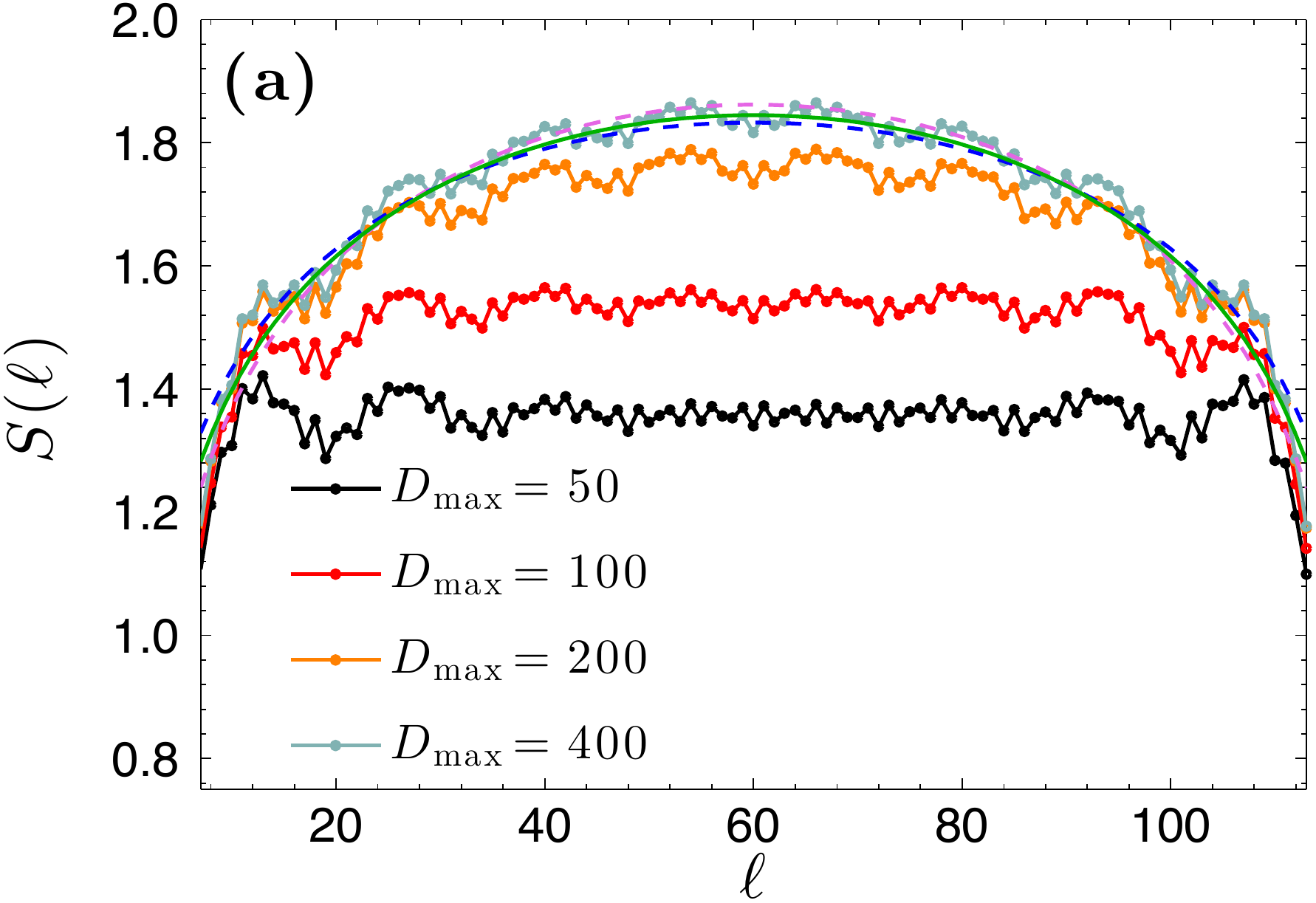}
\hspace{0.4cm}
\includegraphics[width=8cm]{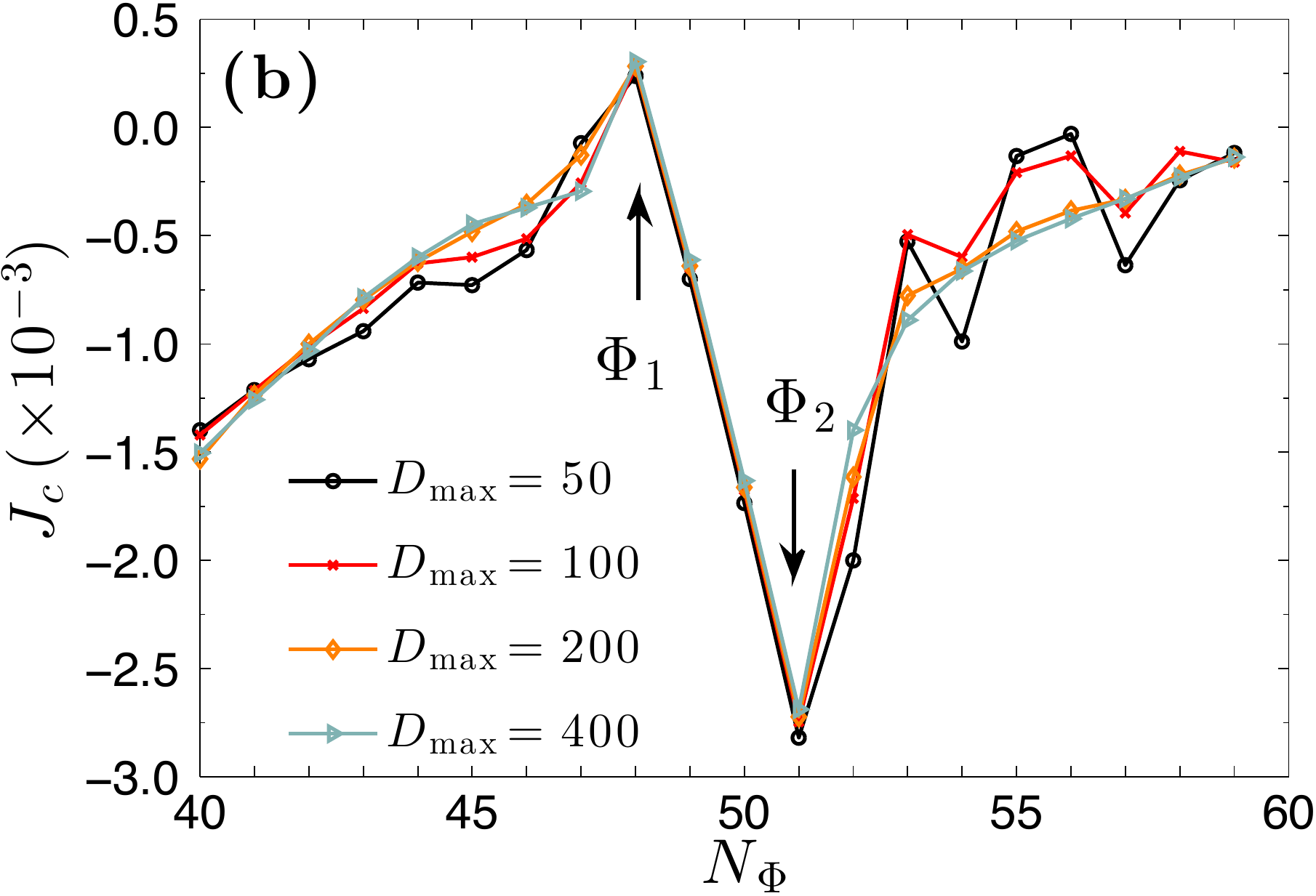}
\caption{(Color online) \textbf{(a)} Numerical data for $S(\ell)$ for different values of $D_{\rm max}$, for $\ell\in[7:113]$ and $N_\Phi=40$. Other simulation parameters as in Fig.~\ref{fig:currentcentralchargeandentanglemententropybosons}. We see that, for $D_{\rm max}=50,100$, the EE in the bulk of the system is flat. We start seeing the behavior predicted by Eq.~\eqref{eq:entanglemententropycalabresecardy} for larger values of $D_{\rm max}$. For $D_{\rm max}=400$, we fit the EE as explained in Sec.~\ref{sec:numericalresults}, and show $S(\ell)$ from Eq.~\eqref{eq:entanglemententropycalabresecardy} with $c_{\rm min}=1.77492$ (blue dashed line), $c_{\rm max}=2.18872$ (purple dashed line), and with the average value $\bar c=(c_{\rm max}+c_{\rm min})/2=1.98182$ (green full line). \textbf{(b)} Numerical data for $J_c$ for different values of $D_{\rm max}$. In contrast to the data on the EE \textbf{(a)}, the chiral current is not drastically affected by the finite value of $D_{\rm max}$. The helical region remains indeed visible, and for $D_{\rm max}=400$ the square-root behavior emerges, for $\Phi>\Phi_2$ (see Fig.~\ref{fig:currentcentralchargeandentanglemententropybosons}).}
\label{fig:currentappendix}
\end{figure*}

\begin{figure}[t]
\centering
\includegraphics[width=8cm]{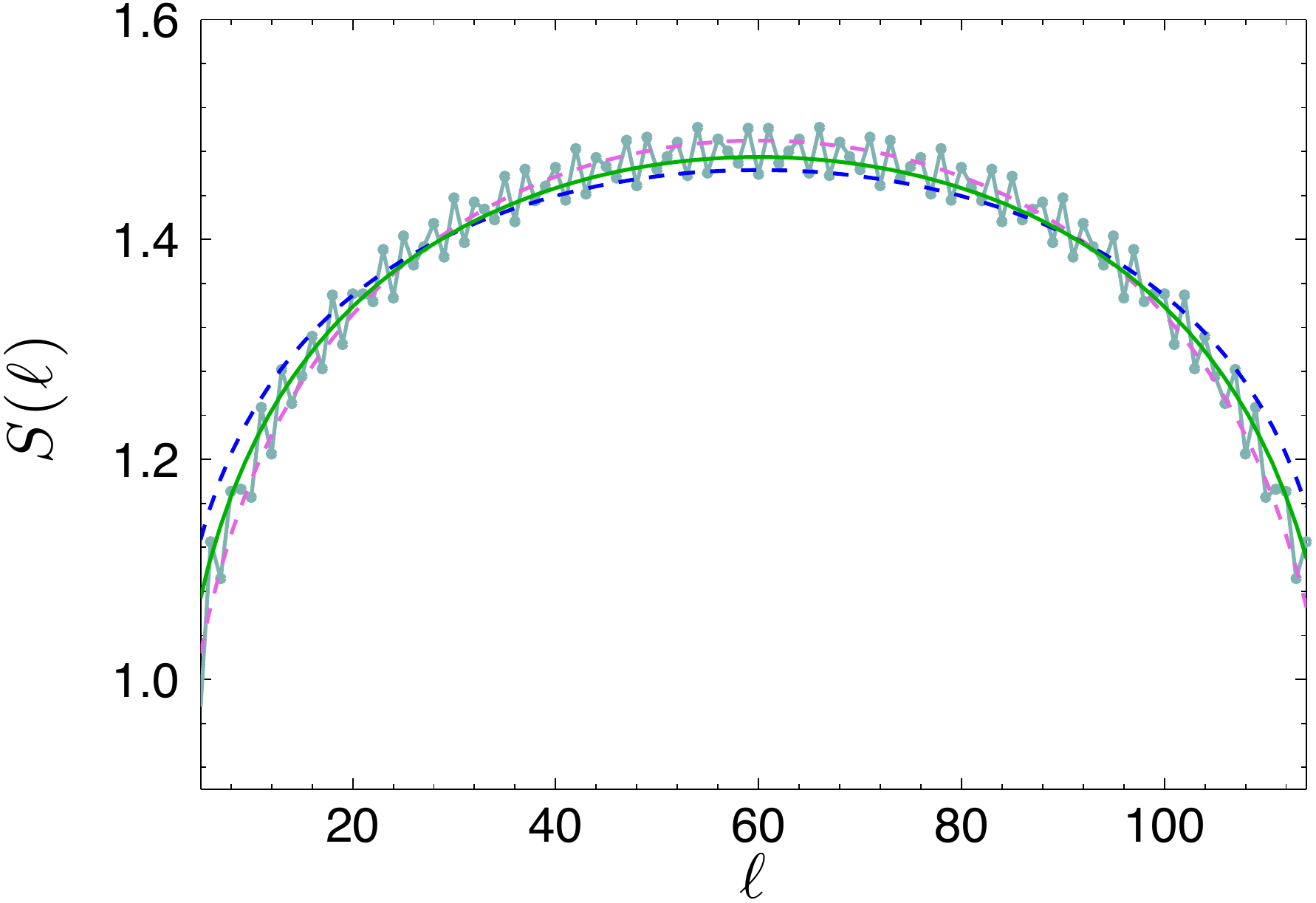}
\caption{(Color online) Numerical data for $S(\ell)$, with $\ell\in[7:113]$, $D_{\rm max}=400$ and $N_\Phi=49$. Other simulation parameters as in Fig.~\ref{fig:currentcentralchargeandentanglemententropybosons}. We fit the EE as explained in Sec.~\ref{sec:numericalresults}, and show $S(\ell)$ from Eq.~\eqref{eq:entanglemententropycalabresecardy} with $c_{\rm min}=0.986867$ (blue dashed line), $c_{\rm max}=1.37001$ (purple dashed line), and with the average value $\bar c=(c_{\rm max}+c_{\rm min})/2=1.178438$ (green full line). The fact that we do not exactly fit $c=1$ can be due to both finite-size effect and to non-perfect convergence of the numerical algorithm.}
\label{fig:eehelicalregion}
\end{figure}

\section{Details on the numerical calculations}
\label{sec:eefit}
In order to highlight the quality of the fit of the EE for the bosonic $\nu=1/2$ Laughlin-like state (for the data in Fig.~\ref{fig:currentcentralchargeandentanglemententropybosons}), here we analyze the numerical results for $N_\Phi=40$ and $D_{\rm max}=400$. We fit the data using Eq.~\eqref{eq:entanglemententropycalabresecardy}, as explained in Sec.~\ref{sec:numericalresults}. We show the result in Fig.~\ref{fig:currentappendix}\textbf{a}, from which it is evident that the leading behavior of the EE, sufficiently away from the C-IC transition point, is given by Eq.~\eqref{eq:entanglemententropycalabresecardy}, and small oscillations are around this leading behavior. The symmetry of the numerical data for $S(\ell)$, as in Fig.~\ref{fig:currentappendix}\textbf{a}, with respect to $\ell=L/2=60$ is an indication of the convergence of the numerical algorithm.

This is not valid, however, as we approach the transition point, where instead large oscillations dominate over the behavior described by Eq.~\eqref{eq:entanglemententropycalabresecardy} (see Fig.~\ref{fig:currentcentralchargeandentanglemententropybosons}\textbf{b}). The deviation from the leading scaling given by Eq.~\eqref{eq:entanglemententropycalabresecardy}, instead, signals the occurrence of the Lifshitz transition.

It is also important to discuss how the EE and $J_c$ are affected by the finite value of $D_{\rm max}$. We thus measure the EE, for a given $N_\Phi$, and $J_c$ for different values of $D_{\rm max}$. From Fig.~\ref{fig:currentappendix}, we see that having a finite value of $D_{\rm max}$ drastically affects the EE, because of the finite amount of states we keep during the variational procedure~\cite{Schollwock201196}: as shown in Fig.~\ref{fig:currentappendix}\textbf{a}, the EE is indeed flat in the bulk of the system (i.e., for $\ell$ sufficiently away from the boundaries) for small $D_{\rm max}$, and tends to be well described by Eq.~\eqref{eq:entanglemententropycalabresecardy}, with $c=2$, for larger values of $D_{\rm max}$.

Instead, the presence of a finite value of $D_{\rm max}$ has a less drastic effect on the chiral current. We show the result in Fig.~\ref{fig:currentappendix}\textbf{b}. The helical region is clearly visible, even for small values of $D_{\rm max}$. The fact that $J_c$ in the non-helical region shows fluctuations for small $D_{\rm max}$ can be regarded as numerical error, because of the large error threshold we have. For $\Phi>\Phi_2$, and $D_{\rm max}=400$, where our data are more precise with respect to those at $D_{\rm max}<400$, the square-root behavior appears, as was shown in Fig.~\ref{fig:currentcentralchargeandentanglemententropybosons}\textbf{a}.

In Fig.~\ref{fig:eehelicalregion}, we show the EE for the data in Fig.~\ref{fig:currentcentralchargeandentanglemententropybosons}, for $N_\Phi=49$ (inside the helical region). Apart from oscillations, the numerical data of the EE well agree with the behavior predicted by Eq.~\eqref{eq:entanglemententropycalabresecardy}. As explained in Sec.~\ref{sec:numericalresults}, the fact that we do not fit $c=1$ can be due to both finite-size effects and to non-perfect convergence of the MPS-based algorithm.

\section{Solution to the exact model}
\label{sec:solutiontotheexactmodel}
In this appendix, we report the solution to the exact model discussed in Sec~\ref{sec:exactmapping}, which was proposed in Ref.~\cite{PhysRevB.92.115446}. We focus on the regime ${U \gg t}$, for general interaction range $\xi$. In this hard-core limit, the interaction becomes a constraint.

The allowed states for $N$ bosonic or fermionic particles on the ladder of length $L$ and OBC are then in one-to-one correspondence with the states of a constrained model. This model consists of $N$ fictitious particles on a ladder of reduced length ${L'=L - (N-1) \xi}$ subject to an additional constraint of not having two fictitious particles on the same rung. Each particle in the reduced lattice corresponds to one particle and $\xi$ empty rungs to its right on the original lattice.

The part ${\hat H_0+\hat H_{\mathrm{ex-int}}}$ takes the shape of a ${\xi' = 0}$ Hamiltonian,
\begin{equation}
\label{nonintmodel}
\hat H'_0+\hat H'_{\mathrm{ex-int}} = \sum_{j}^{L'}\Big[
\hspace{-0.1cm}
-t \hspace{-0.2cm} \sum_{m=\pm1/2} \hspace{-0.2cm} \left(\hat{a}'^\dag_{j,m} \hat a'_{j+1,m}+\mathrm{H.c.}\right)
+U {\hat{n}_j}'^{2} \Big]
\end{equation}
Since site $j$ in the new lattice correspond to location ${j+\xi(\sum_{\ell=1}^{j-1} n'_{\ell})}$ in the original lattice, the inter-chain coupling becomes
\begin{equation}
\label{intHperp}
\hat H_{\perp}' = +t_\perp \sum_{j=1}^{L'} \left(\hat{a'}_{j,-\frac{1}{2}}^\dagger \hat a'_{j,+\frac{1}{2}}e^{-i \Phi [j+\xi(\sum_{\ell=1}^{j-1} n'_{\ell})]}+ \mathrm{H.c.}\right) \,\, ,
\end{equation}
which is nonlocal. However, the nonlocality disappears for a special value of the flux $\Phi=2\pi \kappa/\xi$, where $\kappa$ is a nonnegative integer. Thus, the new particles are subject to the same value of flux ${\Phi'=\Phi}$.

\begin{table}[t]
\centering
\begin{tabular}{|c|c|c|c|c|c|}
\hline
& $\nu$ & $\xi$ & $\Phi$ & $L$ & $n$\\\hline
Physical & $\frac{1}{2\kappa+1}$ & $\geq2\kappa+1$ & $\frac{2\pi\kappa}{\xi}$ & $\left(1+\frac{1}{2\kappa}\right)N\xi$ & $\frac{2\kappa}{(2\kappa+1)\xi}$\\\hline
Remapped & $\frac{N}{2\kappa+N}$ & $0$ & $\frac{2\pi\kappa}{\xi}$ & $\left(1+\frac{N}{2\kappa}\right)\xi$ & $\frac{2\kappa N}{(2\kappa+N)\xi}$\\\hline
\end{tabular}
\caption{Values of the relevant physical quantities of the model, for the physical (true) system, and for the remapped system as explained in Sec.~\ref{sec:exactmapping} (the filling factor for the Laughlin-like state is $\nu=1/p$ with $p=2\kappa+1$).}
\label{tab:mapping}
\end{table}

For a given filling ${\nu}$, the density of the original particles is $n = \nu(\Phi/\pi) =\nu(2 \kappa/\xi)$. The density of the new particles is
\begin{equation}
n'=\frac{N}{L'}=\frac{1}{n^{-1}-\xi\frac{N-1}{N}} \,\, ,
\label{eq:densitydirectmapping}
\end{equation}
and the new filling factor is thus $\nu'=n'\pi/\Phi'={(\nu^{-1} - 2\kappa\frac{N-1}{N})}^{-1}$. The full filling constraint on the density is $N(\xi+1)<L+\xi$, and the integer $\kappa$ is constrained by $\kappa<\frac{1+\xi/L}{2\nu}\frac{\xi}{\xi+1}$. One interesting case where the constraint is satisfied is shown in Table~\ref{tab:mapping}. As an example one may set $\kappa=1$, $\xi=3$, $N=50$ and get
\begin{equation}
\begin{array}{lllll}
\nu\phantom{'}=1/3,&\xi\phantom{'}=3,&\Phi\phantom{'} =2 \pi/3,& L\phantom{'}=225,& n\phantom{'} = 2/9, \\
\nu'=0.96,&\xi'=0,&\Phi'=\Phi,& L'=78, & n' = 0.64
\end{array} \,\, .
\label{eq:simulationmappingvalues}
\end{equation}
Starting from bosons, one can reach similar a mapping by including a Jordan-Wigner transformation. 

The chiral current can also be calculated using the exact mapping in a simpler manner. In the original and reduced models, the chiral current is given by $J_c=-\frac{1}{L}(\partial E_{\rm GS}/\partial\Phi)$ and $J'_c=-\frac{1}{L'}(\partial E'_{\rm GS}/\partial\Phi)$, respectively. As there is a one-to-one correspondence of neighboring states between the models, the GS energies are equivalent. A comparison of these two equations therefore yields the useful relation
\begin{equation}
J_c\{\nu\}=\frac{L'}{L} J'_c\{\nu'\} \,\, .
\label{eq:mappingforthecurrent}
\end{equation}
By setting $\kappa=1$, in the thermodynamic limit ($L\gg\xi$ and $N\gg2\kappa$), we get $J_c\{\nu\}=(1-2\nu)J'_c\{\nu'=\tfrac{\nu}{1-2\nu}\}$, 
constrained by $\nu<\frac{1}{2}\frac{\xi}{\xi+1}$ and $J_c\{n=\frac{1}{\xi+1}\}=\frac{1}{\xi+1}J'_c\{n'=1\}$. This technique allows us to take results from the $\xi'=0$ model and use them to calculate the current at the fractional $\nu=\frac{1}{3}$ model.

\begin{figure}[t]
\centering
\includegraphics[width=8cm]{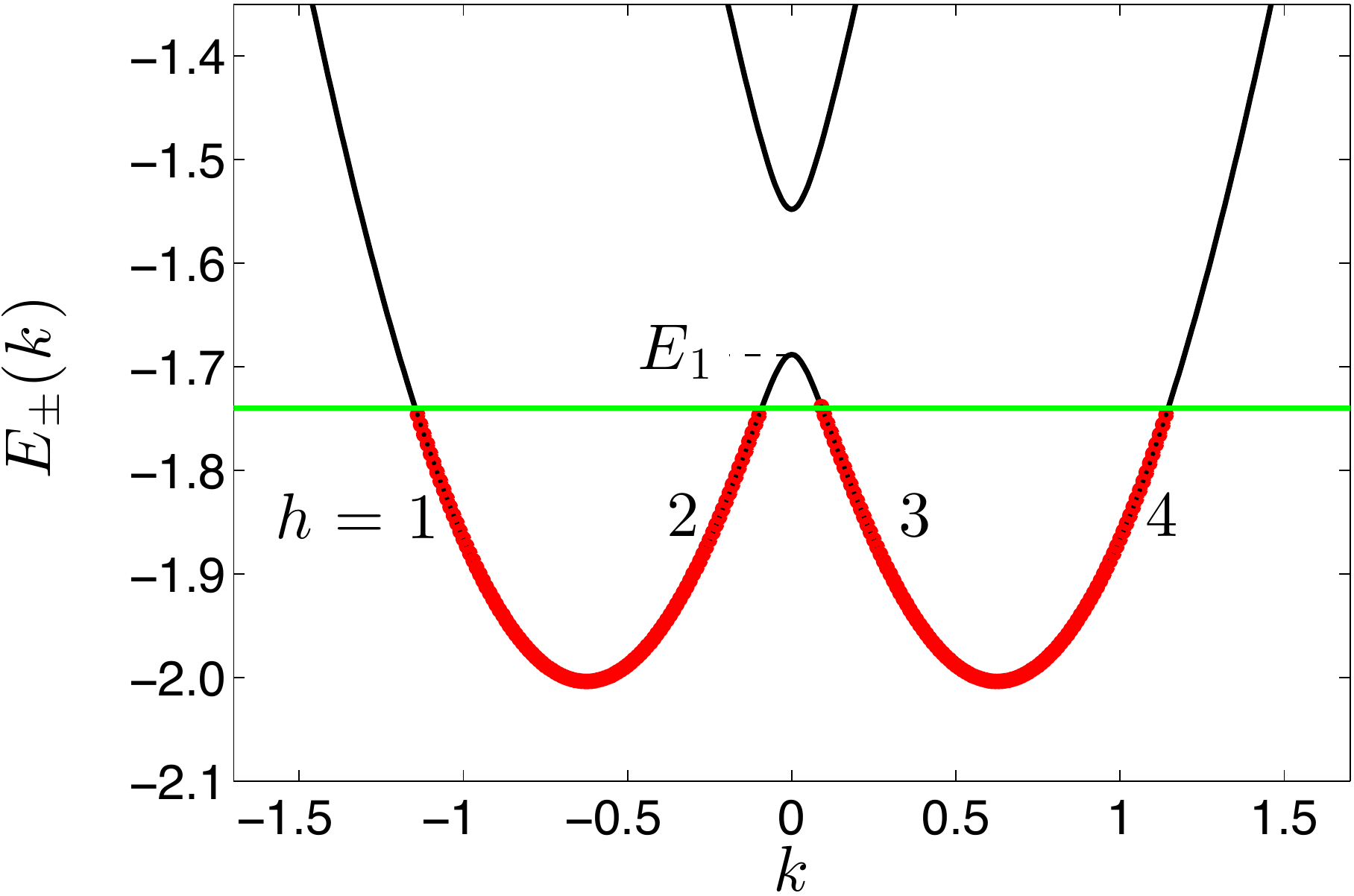}
\caption{(Color online) Band structure $E_\pm(k)$ (solid black lines). Red points represent filled states, and the Fermi energy is represented by the solid green line. The four points ($h=1,2,3,4$), identifying left edge, left bulk, right bulk and right edge respectively, label the four different dispersion relations in Eq.~\eqref{eq:currentcmomentumspaceintegral3}, and $E_1=E_-(0)$.}
\label{fig:currentbandstructureforintegralcalculation}
\end{figure}

\section{Calculation of the current for free fermions}
\label{sec:calculationofthecurrentforfreefermions}
Here, we report the details of the calculation of Eq.~\eqref{eq:currentsingularbehaviorefermienergy}. By denoting the band maximum with $E_1$
(see Fig.~\ref{fig:currentbandstructureforintegralcalculation}), one can split the integral as
\begin{equation}
\left\langle\hat J_m\right\rangle=J_{E_1}-\frac{1}{2\pi}\sum_{h=1}^{4}{(-1)}^h\int_{E_F}^{E_1}dE\,\frac{dk_h}{dE}\,J_m[k_h(E)] \,\, ,
\label{eq:currentcmomentumspaceintegral3}
\end{equation}
where $J_{E_1}$ is the contribution of the lowest band, filled from the band minima to $E_1$. The four functions $k_h (E)$ represent the band dispersion relations around the four Fermi points $h=1,2,3,4$ (see Fig.~\ref{fig:currentbandstructureforintegralcalculation}). For simplicity, we rename $j_h=\int_{E_F}^{E_1}dE\,\frac{dk_h}{dE}\,J_m[k_h(E)]$. We are interested in describing the dependence $\langle\hat J_m\rangle(E_F)$ as $E_F\rightarrow E_1$. For simplicity, we show the calculation for $m=+1/2$.

We first observe that, for the modes $h=1,4$, the dispersion relation is almost linear in $E$, i.e. $k_{1,4}(E)\simeq\mp\left(\frac{E-E_1}{v_F}+k_{e,0}\right)$, where $k_{e,0}$ is a proper momentum offset; thus $dk_{1,4}/dE=\mp1/v_F$. The functions $j_{1,4}$ do not display any singular behavior because the integrand is analytic. For sufficiently small $t_\perp$, one can approximate the $h=1$ mode to be polarized $m=-1/2$ and the $h=4$ one to be polarized $m = +1/2$. Thus, $j_1 \sim 0$ and
\begin{eqnarray}
j_4&\simeq&\frac{2t}{v_F}\cos(k_{e,0}-\Phi/2)\int_{E_F}^{E_1}dE\,\sin\left(\frac{E-E_1}{v_F}\right)\nonumber\\
\nonumber\\
&&+\frac{2t}{v_F}\sin(k_{e,0}-\Phi/2)\int_{E_F}^{E_1}dE\,\cos\left(\frac{E-E_1}{v_F}\right) \,\, . \nonumber\\
\label{eq:jfourterm}
\end{eqnarray}
In the limit $E_F\rightarrow E_1$, to the first order in $E_1-E_F$, the second term in Eq.~\eqref{eq:jfourterm} yields the linear dependence $j_4\simeq x_4(E_1-E_F)$, where $x_4=\frac{2t}{v_F}\sin(k_{e,0}-\Phi/2)$.

The modes $h=2,3$ are treated similarly. The lower band around $k=0$ can be approximated as $E_-(k)\simeq E_1-ak^2/(|k|+b)$, for some real and positive $a$ and $b$. Since $k_2(E)=-k_3(E)$, by renaming for simplicity $k_3(E)=k_b(E)$, one has $dk_{b}/dE=-\frac{1}{2a}-\frac{1}{2}\sqrt{\frac{b}{a}}{(E_1-E)}^{-1/2}$. The current around $k=0$ can be approximated as $J_m(k)\simeq\left(\frac{1}{2}+\alpha k\right)\sin(k-\Phi/2)$, for $\alpha>0$. Recalling Eq.~\eqref{eq:currentcmomentumspaceintegral3}, we obtain:
\begin{equation}
j_3-j_2=\int_{E_F}^{E_1}dE\,\frac{dk_b}{dE}\,\left\{J_m[k_b(E)]+J_m[-k_b(E)]\right\} \,\, .
\end{equation}
One can see, by explicit inspection, that $J_m[k_b(E)]+J_m[-k_b(E)]$ has no singular behavior in the limit $E_F\rightarrow~ E_1$. A singular behavior arises from the divergence in $dk_b/dE$, which leads to $j_3-j_2=x_2(E_1-E_F)+A_1\sqrt{E_1-E_F}$, for some coefficients $x_2$ and $A_1$.

In the helical region, slightly above the $c=2\to c=1$ transition, only helical edge modes are found, with linear dispersion relation. Their contribution to the current is found exactly in the same way, and gives $\langle\hat J_m\rangle=A_3(E_F-E_1)$, for some coefficient $A_3$. We thus find Eq.~\eqref{eq:currentsingularbehaviorefermienergy}, for some coefficients $A_1$, $A_2=x_4+x_2$, and $A_3$.

The same framework can be used to describe the behavior of the current for the $c=1\to c=2$ transition, i.e. when the Fermi energy is slightly above the minimum of the upper band, $E_2\equiv E_+(0)$ (see Fig~\ref{fig:currentbandstructureforintegral}). In this case, the current, as a function of the Fermi energy, is expected to vary as $\langle\hat J_m\rangle=J'_{E_2}+A_4\sqrt{E_F-E_2}+A_5(E_F-E_2)$, for some $J'_{E_2}$, $A_4$ and $A_5$.

\end{document}